\documentclass[aps,twocolumn,pra,superscriptaddress]{revtex4}
\usepackage{epsfig,graphicx,times}
\usepackage{amstext}
\usepackage{amsmath}
\usepackage{mathrsfs}
\usepackage{amssymb}
\usepackage{graphicx}
\usepackage{latexsym}
\usepackage{bm}
\usepackage{float}
\usepackage[colorlinks,citecolor=blue,linkcolor=blue,hyperindex]{hyperref}
\usepackage{orcidlink}

\begin{document}

\title{Coupling Enhancement and Symmetrization in Dissipative Optomechanical Systems}
\author{Cheng Shang \orcidlink{0000-0001-8393-2329} \footnote{c-shang@iis.u-tokyo.ac.jp}}
\affiliation{Department of Physics, The University of Tokyo, 5-1-5 Kashiwanoha, Kashiwa, Chiba 277-8574, Japan}
\affiliation{Analytical Quantum Complexity RIKEN Hakubi Research Team, RIKEN Center for Quantum Computing (RQC), Wako, Saitama 351-0198, Japan}
\author{H. Z. Shen \orcidlink{0000-0002-4017-7367} \footnote{shenhz458@nenu.edu.cn}}
\affiliation{Center for Quantum Sciences and School of Physics, Northeast Normal University, Changchun 130024, China}

\begin{abstract}
Observing few-photon optomechanical effects remains a significant challenge in optomechanical systems. To investigate intrinsic radiation-pressure-induced nonlinear effects in the few-photon regime, it is essential to strengthen the interaction between few photons and a finite number of phonons. In this work, we enhance the radiation-pressure nonlinearity by introducing a two-laser coherent driving scheme together with an enhanced cross-Kerr nonlinearity, resulting in a setup that can be effectively described within a circuit QED platform. By properly tuning the two driving laser fields and the cross-Kerr interaction so that the effective optomechanical coupling becomes real, we theoretically establish a symmetric optomechanical model in which the photon and phonon modes exhibit analogous fluctuation dynamics. Within this framework, we analyze the optimal reciprocal transport of the input laser field and identify the critical boundary associated with the onset of different coupling regimes. We also compare the optical signal scattering behavior in both dissipative equilibrium and nonequilibrium symmetric optomechanical systems, with and without non-rotating-wave contributions. Our work provides a controllable route to enhance optomechanical coupling, extending into the ultrastrong-coupling regime, and opens opportunities for exploring few-photon optomechanical effects.
\end{abstract}

\maketitle

\section{Introduction}
Cavity optomechanics is a rapidly developing research field involving quantum optics and nanoscience \cite{ref-1,ref-2,ref-3}. Work in this field focuses on studying the radiation-pressure (RP) coupling between electromagnetic and mechanical degrees of freedom \cite{ref-4,ref-5,ref-6}. Recently, particular interest has been devoted to exploring optomechanical effects in the few-photon regime \cite{ref-7,ref-8,ref-9}, as exploiting the intrinsic nonlinear RP interaction (\textit{note that, as the dominant contribution to optomechanical nonlinearity, it is commonly referred to as the optomechanical coupling}) of the cavity optomechanics would open the doors to a much vaster realm of possibilities in the quantum regime. Many unusual phenomena appear in this regime, such as the appearance of phonon sidebands in the cavity emission spectrum \cite{ref-10}, the photon blockade effect due to a weakly driving laser \cite{ref-11,ref-12,ref-13,ref-14}, and the generation of macroscopic quantum superposition \cite{ref-15,ref-16}.

However, optomechanical effects in the few-photon regime, especially at the single-photon level, have not yet been observed under existing experimental techniques \cite{ref-2}. This difficulty arises because the radiation-pressure coupling involving only a few photons is typically too weak to be resolved from environmental noise in the standard optomechanical systems \cite{ref-17}. Consequently, enhancing the optomechanical coupling to reach the strong- or even ultrastrong-coupling regime in cavity optomechanics involving few photons remains a nontrivial task. To date, many schemes have been proposed to enhance the optomechanical coupling in the few-photon regime. These schemes include the generation of collective density excitation of the Bose-Einstein condensate \cite{ref-18}, the construction of an array of mechanical resonators \cite{ref-19}, the usage of the squeezing cavity mode \cite{ref-20,ref-21}, the exploitation of mechanical amplification \cite{ref-22}, the application of delayed quantum feedback \cite{ref-23}, and the utilization of the critical property of the lower-branch polariton cavity coupling \cite{ref-24}. In addition, recently, an all-optical system based on a Fredkin-type interaction has been proposed to simulate the ultrastrong optomechanical coupling \cite{ref-25}. Yet, there are two common issues in these schemes \cite{ref-26}. One is how to guarantee that the system operates in the few-photon regime. The other is how to ensure that radiation-pressure-induced nonlinear effects can be observed in an optomechanical system if additional sources of nonlinearity are introduced.

Motivated to resolve these issues, in this paper, we focus on a novel strategy to enhance the nonlinear effects, enabling access to the strong and even ultrastrong coupling regime with few photons in a circuit quantum electrodynamics (QED) platform.~First, we propose a method to achieve controllable enhancement of the optomechanical coupling in the few-photon regime, which can be simulated on a circuit QED platform.~In contrast to the hybrid cavity optomechanics or optomechanical-like systems, we stress that our scheme is realized by considering only two points. One contribution stems from the enhanced second-order nonlinearity of the cavity resonance frequency beyond the radiation-pressure interaction, known as the cross-Kerr (CK) interaction \cite{ref-27,ref-28}, which is an inherent nonlinear interaction accompanying optomechanical coupling \cite{ref-29,ref-30,ref-31}. The other is to make sure that the mechanical resonator and the optical cavity are coherently driven by an appropriate high-power laser \cite{ref-32,ref-33} and a low-power one \cite{ref-34}, respectively.~Second, by adjusting the parameters of the two driving lasers and the CK interaction so that the effective optomechanical coupling may become a real number, we theoretically propose effective symmetric optomechanical dynamics in the few-photon regime, in which the quantum fluctuation dynamics of the photon and phonon modes exhibit analogous forms.~Third, we control the optomechanical coupling strength involving a few photons and a finite number of phonons via dual coherent laser driving and the enhanced cross-Kerr nonlinearity, achieving sequential transitions from the weak-coupling to the ultrastrong-coupling regime \cite{ref-35,ref-36}. By observing the critical behavior of the optimal transmission of the laser field, we identify the boundary point of the optomechanical strong coupling. In addition, we study the optimal reciprocal transport in symmetric optomechanical dynamics \cite{ref-37,ref-38,ref-39}. We also compare the scattering behavior of the laser field when the decay rate of the cavity field matches the damping rate of the mechanical oscillator or not before and after the rotating-wave approximation (RWA).

The remainder of this paper is organized as follows. In Sec.~\ref{section2}, we introduce the physical model and derive the dynamical equations governing the system. In Sec.~\ref{section3}, an effective symmetric optomechanical Hamiltonian is developed in the few-photon regime, and the corresponding optical and mechanical driving powers are estimated. In Sec.~\ref{section4}, we analyze the transmission behavior of the laser field in the symmetric optomechanical system, both with and without the rotating-wave approximation. Finally, in Sec.~\ref{section6}, we present our conclusions and outlook.

\section{Coupling control of optomechanical systems}\label{section2}

\subsection{Derivation of controllable coupling setup}
\begin{figure}[h]
\centering
\includegraphics[angle=0,width=0.478\textwidth]{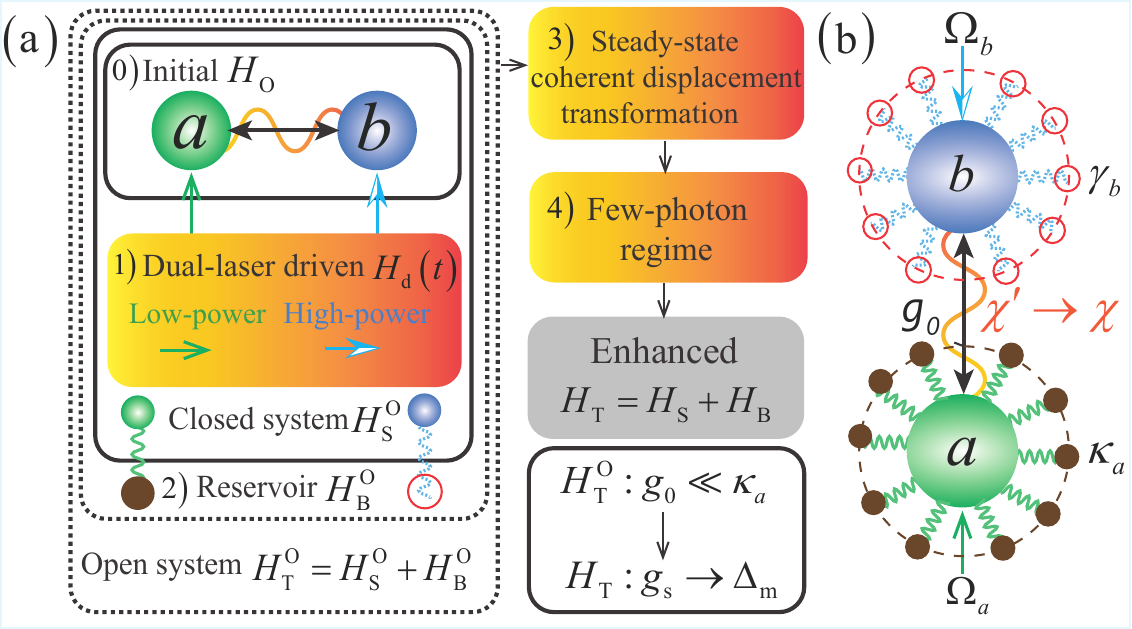}
\caption{$\left( \rm{a} \right)$ A flowchart for obtaining an enhanced Hamiltonian ${H_{\rm{T}}}$. Initially, we consider an original (i.e., at single-photon level) optomechanical Hamiltonian, denoted by ${H_{\rm{O}}}$. The first step is to introduce two laser beams ${H_{\rm{d}}}\left( t \right)$ to drive the initial system. Next, we extend the above closed system $H_{\rm{S}}^{\rm{O}}$ to an open quantum system $H_{\rm{T}}^{\rm{O}}$ by adding an environmental Hamiltonian $H_{\rm{B}}^{\rm{O}}$. Then, we perform a steady-state coherent displacement transformation on it. As the last step, we safely neglect the originally enhanced CK interaction term to obtain an enhanced optomechanical Hamiltonian ${H_{\rm{T}}}$ in the few-photon regime. $\left( \rm{b} \right)$ Schematic illustration of an open optomechanical setup $H_{\rm{T}}^{\rm{O}}$ [left side of $\left( \rm{a} \right)$], which is composed of a single-mode optical field $a$ (lower side) and a mechanical mode $b$ (upper side). The optical mode is coupled to the mechanical mode via both the radiation-pressure (RP) interaction ${g_{\rm{0}}}$ $\left(  \leftrightarrow  \right)$ and the enhanced CK interaction $\chi $ $\left(  \sim  \right)$. The optical mode is coherently driven by a low-power laser $\Omega_a$, thereby operating in the few-photon regime, while the mechanical mode is coherently driven by a high-power laser $\Omega_b$, leading to the excitation of a finite number of phonons. As a result of the dual coherent laser driving, the single-photon optomechanical coupling $g_0$ is enhanced to an effective optomechanical coupling $g_s$ in the few-photon regime. Additionally, the near-resonant condition between the system modes and the corresponding reservoir modes is also considered in this open quantum system.}\label{Fig-1}
\end{figure}

\begin{figure}[t]
\centering
\includegraphics[angle=0,width=0.4\textwidth]{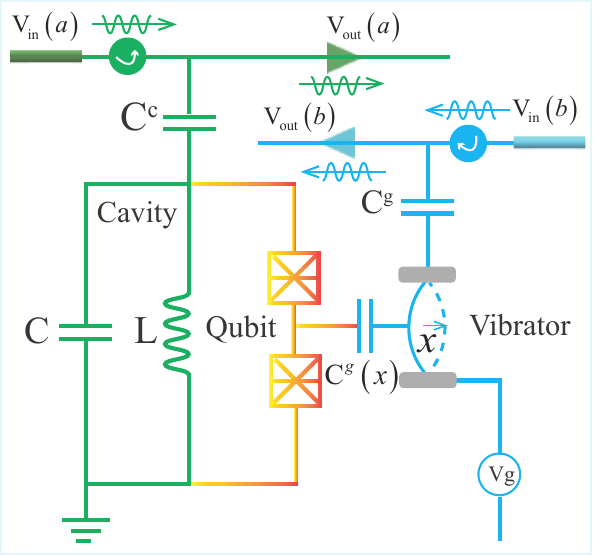}
\caption{The CK enhanced setup based on a circuit quantum electrodynamics platform, which provides an effective description of the enhanced optomechanical coupling. This setup is made of superconducting qubits (orange) coupled to an on-chip cavity (green) and a nano-mechanical resonator (blue).} \label{Fig-add}
\end{figure}

For the standard optomechanical setup, the strength of the single-photon radiation-pressure (RP) nonlinear interaction (\textit{RP nonlinearity as the dominant optomechanical nonlinearity, it is commonly referred to as the optomechanical coupling}) ${g_{\rm{0}}}$ is far less than the decay rate of the optical cavity ${\kappa _a}$. Figure~\ref{Fig-1}(a) presents the procedure for driving an optomechanical system, effectively described by a circuit quantum electrodynamics (QED), into the ultrastrong coupling regime characterized by the coupling strength $g_{\rm{s}}$ through laser control in an open quantum system. Here, ultrastrong coupling denotes the regime in which the enhanced optomechanical interaction $g_{\rm{s}}$ becomes a considerable fraction of the detuning of the mechanical resonance frequency ${{\Delta _{\rm{m}}}}$ \cite{ref-40}.

We first initialize the system by introducing a standard optomechanical Hamiltonian describing a single-mode optical cavity field coupled to a single-mode mechanical oscillator [see Fig.~\ref{Fig-1}(b)]. In particular, we consider the lowest-order correction to the optomechanical interaction beyond the radiation-pressure (RP) term and neglect the rapidly oscillating terms, $\hbar \chi {a^\dag }a{b^\dag }{b^\dag }$ and $\hbar \chi {a^\dag }abb$, see Appendix \ref{appendix A} for details \cite{ref-41,ref-42,ref-43,ref-44,ref-79,ref-80}. Then, the Hamiltonian of the standard optomechanical model reads
\begin{eqnarray}
H_{\rm{O}} = \hbar {\omega _{\rm{c}}}{a^\dag }a \!+\! \hbar {\omega _{\rm{m}}}{b^\dag }b \!-\! \hbar {g_{\rm{0}}}{a^\dag }a\left( {{b^\dag } + b} \right){\rm{ \!+ }}\hbar \chi ' {a^\dag }a{b^\dag }b,\label{eq-1}
\end{eqnarray}
where ${a^\dag }$ $\left( a \right)$ and ${b^\dag }$ $\left( b \right)$ are the creation (annihilation) operators of the optical mode and the mechanical mode with the corresponding resonance frequencies ${\omega _{\rm{c}}}$ and ${\omega _{\rm{m}}}$, respectively. Here, the first and second terms in Eq.~(\ref{eq-1}) represent the free parts of the optical and mechanical modes, respectively. The third term in Eq.~(\ref{eq-1}) describes the original RP coupling between the two modes, where $g_0$ denotes the optomechanical coupling strength at the single-photon level, quantifying the interaction between a single photon and a single phonon. The fourth quadratic term in Eq.~(\ref{eq-1}) represents the original CK interaction between the two modes, with coupling strength ${\chi '}$.

Although the RP coupling is typically the leading interaction in optomechanical systems, the second-order nonlinear term can be substantially enhanced via Josephson junctions \cite{ref-45} or superconducting qubits \cite{ref-46} in optical implementations, a fact that has been experimentally verified \cite{ref-88,ref-89,ref-90,ref-91}. In our study, we focus on a CK-enhanced setup in which the CK nonlinear interaction becomes much stronger than the original RP nonlinearity \cite{ref-47,ref-48,ref-49}. Under this experimentally accessible setting within a circuit QED platform as shown in Fig.~\ref{Fig-add}, the third term in Eq.~(\ref{eq-1}) becomes negligible compared to the enhanced CK term, allowing us to safely adopt the approximate Hamiltonian $H_{{\rm{O}}}^{{\rm{app}}} = \hbar {\omega _{\rm{c}}}{a^\dag }a + \hbar {\omega _{\rm{m}}}{b^\dag }b + \hbar \chi {a^\dag }a{b^\dag }b$. This Hamiltonian conserves the photon number due to the commutation relation $[ {{a^\dag }a,H_{\rm{{O}}}}] = 0$ \cite{ref-Lu053703}. In the following, based on the CK-enhanced circuit QED setup that effectively describes the enhanced optomechanical Hamiltonian~\cite{ref-50,ref-52}, we present four steps to obtain a controllable optomechanical coupling model.

The first step.~In order to enhance the optomechanical coupling, we introduce dual-laser coherent driving to the system described by the approximate Hamiltonian $H_{{\rm{O}}}^{{\rm{app}}}$. The driving Hamiltonian is given by
\begin{eqnarray}
{H_{\rm{d}}}\left( t \right) = \hbar \left( {\Omega _a^*a{e^{i{\omega _{{L_a}}}t}} + \Omega _b^*b{e^{i{\omega _{{L_b}}}t}}} \right) + {\rm{H.c.}},\label{eq-2}
\end{eqnarray}
where the optical mode is coherently driven by a weak monochromatic field (thereby operating in the few-photon regime) with ${{\omega _{{L_a}}}}$ and ${\Omega _a}$ being the driving frequency and complex amplitude, respectively. The mechanical mode, on the other hand, is coherently driven by a strong monochromatic field (leading to the excitation of a finite number of phonons), where ${{\omega _{{L_b}}}}$ and ${\Omega _b}$ are the driving frequency and complex amplitude, respectively. The total Hamiltonian of the closed system is now given by $H_{\rm{S}}^{\rm{O}}\left( t \right) = H_{{\rm{O}}}^{{\rm{app}}} + {H_{\rm{d}}}\left( t \right)$. To make the Hamiltonian independent of time, we then move to the rotating frame of the frequency, which makes the Hamiltonian of the closed system as follows:
\begin{eqnarray}
{H_{\rm{S}}^{\rm{O}}} &=& +U\left( t \right)H_{\rm{S}}^{\rm{O}}\left( t \right){U^\dag }\left( t \right) - iU\left( t \right){{\dot U}^\dag }\left( t \right)\nonumber\\
 &=& +\hbar {\Delta _{\rm{c}}}{a^\dag }a + \hbar {\Delta _{\rm{m}}}{b^\dag }b {\rm{ + }}\hbar \chi {a^\dag }a{b^\dag }b \nonumber\\
 &&+\hbar \left( {\Omega _a^*a + \Omega _b^*b + {\rm{H.c.}}} \right),\label{eq-3}
\end{eqnarray}
where we used the unitary transformation of the form $U\left( t \right) = \exp [ {i( {{\omega _{{L_a}}}{a^\dag }a + {\omega _{{L_b}}}{b^\dag }b})t} ]$. The parameter ${\Delta _{\rm{c}}} = {\omega _{\rm{c}}} - {\omega _{{L_a}}}$ is the detuning of the resonant frequency ${\omega _{\rm{c}}}$ of the optical cavity $a$ from the driving frequency ${{\omega _{{L_a}}}}$, while ${\Delta _{\rm{m}}} = {\omega _{\rm{m}}} - {\omega _{{L_b}}}$ is the detuning of ${\omega _{\rm{m}}}$ of the mechanical oscillator $b$ from the driving frequency ${{\omega _{{L_b}}}}$.

The second step.~In order to extend the system to an open quantum one, we further add a system-reservoir interaction using a quantum operator approach \cite{ref-53}. The total Hamiltonian of this field-reservoir system is now written as $H_{\rm{T}}^{\rm{O}} = H_{\rm{S}}^{\rm{O}} + H_{\rm{B}}^{\rm{O}}$ with
\begin{eqnarray}
H_{\rm{B}}^{\rm{O}} \!\!&=\!\!& + \sum\limits_{\rm{k}} {\hbar \left[ {{\omega _{a{\rm{k}}}}\Gamma _{a{\rm{k}}}^\dag {\Gamma _{a{\rm{k}}}} + {g_{a{\rm{k}}}}\left( {{a^\dag }{\Gamma _{a{\rm{k}}}} + a\Gamma _{a{\rm{k}}}^\dag } \right)} \right]}  \nonumber\\&& +\! \sum\limits_{\rm{k}} {\hbar \left[ {{\omega _{b{\rm{k}}}}\Gamma _{b{\rm{k}}}^\dag {\Gamma _{b{\rm{k}}}} + {g_{b{\rm{k}}}}\left( {{b^\dag }{\Gamma _{b{\rm{k}}}} + b\Gamma _{b{\rm{k}}}^\dag } \right)} \right]},
\label{eq-4}
\end{eqnarray}
where ${\Gamma _{a{\rm{k}}}^\dag }$ $\left( {{\Gamma _{a{\rm{k}}}}} \right)$ and ${\Gamma _{b{\rm{k}}}^\dag }$ $\left( {{\Gamma _{b{\rm{k}}}}} \right)$ are, respectively, the creation (annihilation) operators of the reservoirs for the optical mode and the mechanical mode. The present reservoir consists of many harmonic oscillators with closely spaced frequencies ${{\omega _{a{\rm{k}}}}}$ $\left( {{\omega _{b{\rm{k}}}}} \right)$. The ${{g_{a{\rm{k}}}}}$ and ${{g_{b{\rm{k}}}}}$ terms represent the corresponding system-reservoir interactions.

The third step.~To achieve steady-state optomechanical coupling enhancement in an open optomechanical system, we modulate the detuning of the resonant frequency of the optical cavity ${\Delta _c}$ by driving the mechanical oscillator $b$ (the movable mirror) with appropriate high-power laser ${{\Omega _b}}$, thereby enhancing the optomechanical coupling into the ultrastrong-coupling regime; that is, when the mechanical oscillator is coherently driven by a specific intense monochromatic laser ${\Omega _b}$ that satisfies $\left| {{\Omega _b}} \right| \gg \left| {{\Omega _a}} \right|$, the excitation number in the mechanical mode $b$ is large, and then $b$ contains a steady-state coherent part $\beta_s $.~In other words, we employ the coherent-displacement transformation \cite{ref-54,ref-55} of the mechanical mode of the form $D\left( \beta  \right)b{D^\dag }\left( \beta  \right) = b - \beta $ and ${D\left( \beta  \right){b^\dag }{D^\dag }\left( \beta  \right) = b^\dag - {\beta ^*}}$, where the unitary displacement operator is given by ${D_b}\left( \beta  \right) = \exp \left( {\beta {b^\dag } - {\beta ^*}b} \right)$ in the coherent-state representation. Since our purpose here is to obtain an optomechanical coupling enhancement at a steady state, in the following, we focus on the steady-state displacement solution ${\beta _s}$; we assume that the timescale of the system approaching its steady state is much shorter than other evolution timescales~\cite{ref-56}. The specific form of ${\beta _s}$ will be given by the dynamics equation later. The total Hamiltonian $H_{\rm{T}}^{\rm{O}}$ after the transformation can be written as $H_{\rm{T}}^{{\rm{tra}}} = H_{\rm{S}} + {H_{\rm{B}}}$ with
\begin{eqnarray}
H_{\rm{S}} &=& \!\hbar \Delta _{\rm{c}}^0{a^\dag }a \!+\! \hbar \left( {{\Delta _{\rm{m}}} \!+\! \chi {a^\dag }a} \right){b^\dag }b \!-\! \hbar \chi {a^\dag }a\left( {{\beta _{\rm{s}}}{b^\dag } \!+\! \beta _{\rm{s}}^*b} \right) \nonumber\\
 &&+ \hbar \left( {\Omega _a^*a + \Omega _b^*b - \beta _{\rm{s}}^*{\Delta _{\rm{m}}}b + {\rm{H}}.{\rm{c}}.} \right) \nonumber\\&&+ \hbar \Delta _{\rm{m}}\beta _{\rm{s}}^*{\beta _{\rm{s}}} - \hbar \Omega _b^*{\beta _{\rm{s}}} - \hbar {\Omega _b}\beta _{\rm{s}}^*,\label{eq-5}\\
{H_{\rm{B}}} &=& H_{\rm{B}}^{\rm{O}} - \hbar \sum\limits_{\rm{k}} {{g_{b{\rm{k}}}}} \left( {\beta _{\rm{s}}^*{\Gamma _{b{\rm{k}}}} + {\beta _{\rm{s}}}\Gamma _{b{\rm{k}}}^\dag } \right),\label{eq-6}
\end{eqnarray}
where $\Delta _{\rm{c}}^0 = {\Delta _{\rm{c}}} + \chi \beta _{\rm{s}}^*{\beta _{\rm{s}}}$ is the normalized optical cavity detuning including the frequency shift caused by the high-power laser coherent driving.

The fourth step.~Since ${\beta _s}$ is controlled by the mechanical laser coherent driving, the RP optomechanical coupling can be enhanced to an effective strength $\chi {\beta _{\rm{s}}}$. After performing the coherent-displacement transformation at the steady state, the optomechanical interaction takes the form $\hbar {a^\dag }a\left( {{g_s}{b^\dag } + g_s^*b} \right)$, where ${g_{\rm{s}}} = \chi {\beta _{\rm{s}}}$. Under experimentally accessible conditions in a circuit QED platform in the few-photon regime, the parameters of interest satisfy the condition $\chi {\beta _s} \sim {\Delta _{\rm{m}}} \gg \chi $ with ${\chi  \mathord{\left/{\vphantom {\chi  {{\Delta _{\rm{m}}}}}} \right. \kern-\nulldelimiterspace} {{\Delta _{\rm{m}}}}} \approx {10^{ - 3}}$ \cite{ref-56}.~Under this condition, the enhanced cross-Kerr interaction term $\chi {a^\dag }a{b^\dag }b$ can be negligible compared with both the mechanical detuning term ${\Delta _{\rm{m}}}{b^\dag }b$ and the enhanced optomechanical interaction term $\chi {a^\dag }a\left( {{\beta _s}{b^\dag } + \beta _s^*b} \right)$. Therefore, the enhanced cross-Kerr interaction term $\chi {a^\dag }a{b^\dag }b$ in Eq.~(\ref{eq-5}) can be safely neglected \cite{ref-56}, yielding the final enhanced optomechanical Hamiltonian ${H_{\rm{T}}}$ (after shifting the origin of the energy) as follows:
\begin{eqnarray}
{H_{\rm{T}}} &=& +\hbar \Delta _{\rm{c}}^0{a^\dag }a + \hbar {\Delta _{\rm{m}}}{b^\dag }b - \hbar {a^\dag }a\left( {{g_{\rm{s}}}{b^\dag } + g_{\rm{s}}^*b} \right) \nonumber\\&&+ \hbar \left( {\Omega _a^*a + \Omega _b^*b - \beta _{\rm{s}}^*{\Delta _{\rm{m}}}b + {\rm{H}}.{\rm{c}}.} \right) + {H_{\rm{B}}}.\label{eq-7}
\end{eqnarray}
Thus, the effective optomechanical coupling is enhanced from $g_0$, which describes the coupling between a single photon and a single phonon in Eq.~(\ref{eq-1}), to $g_s$, corresponding to the coupling involving a few photons and a finite number of phonons in Eq.~(\ref{eq-7}).

\subsection{Nonlinear quantum Langevin equations}
To introduce quantum damping and noise in Eq.~(\ref{eq-7}) and to give the specific form of ${\beta _s}$, we now describe the coupling between the system and the reservoir in terms of input-output formalism \cite{ref-57}. This formalism provides us with equations of motion for the amplitude of the optical cavity field $a$ and analogously for the mechanical amplitude $b$. Substituting the final enhanced optomechanical Hamiltonian ${H_{\rm{T}}}$ into the Heisenberg equation and taking the dissipation terms with ${\kappa _a}$ and ${\gamma _b}$, as well as the corresponding noise terms with ${a_{{\rm{in}}}}$ and ${b_{{\rm{in}}}}$ into account, we find a set of closed integro-differential equations for the operators of the optical mode and mechanical mode as follows
\begin{eqnarray}
\!\!\!\! \dot a \! &=& \!\! - \left[ {i\Delta _{\rm{c}}^0 + \frac{{{\kappa _a}}}{2} - i\left( {{g_{\rm{s}}}{b^\dag } + g_{\rm{s}}^*b} \right)} \right]a  - i{\Omega _a} + \sqrt {{\kappa _a}} {a_{{\rm{in}}}},\label{eq-8}\\
\!\!\!\! \dot b \! &=&  \!\! - \left( {i{\Delta _{\rm{m}}} + \frac{{{\gamma _b}}}{2}} \right)b + i{g_{\rm{s}}}{a^\dag }a + \sqrt {{\gamma _b}} {b_{{\rm{in}}}} + {{\dot \beta }_{\rm{s}}},\label{eq-9}
\end{eqnarray}
where ${{\kappa _a}}$ is the decay rate of the optical cavity field, ${{\gamma _b}}$ is the mechanical damping rate.

The time derivative of the steady-state displacement amplitude ${{\dot \beta }_{\rm{s}}}$ is equal to zero. See Appendix \ref{appendix B} for a detailed derivation of the Heisenberg-Langevin equation \cite{ref-58,ref-59}.~From the results in Appendix \ref{appendix B}, we can see that the condition for the Heisenberg-Langevin equation to remain unchanged after the coherent displacement transformation is $\dot \beta \left( t \right) + \left( {i{\Delta _{\rm{m}}} + 0.5{\gamma _b}} \right)\beta \left( t \right) - i{\Omega _b} = 0$. The solution gives the transient displacement amplitude $\beta \left( t \right)$ depending on the mechanical laser drive ${\Omega _b}$ in the form $\beta \left( t \right) = {{{\Omega _b}} \mathord{\left/{\vphantom {{{\Omega _b}} {\left( {{\Delta _{\rm{m}}} - i0.5{\gamma _b}} \right)}}} \right.\kern-\nulldelimiterspace} {\left( {{\Delta _{\rm{m}}} - i0.5{\gamma _b}} \right)}} + C\left( {{t_0}} \right)\exp \left[ { - \left( {i{\Delta _{\rm{m}}} + 0.5{\gamma _b}} \right)t} \right]$. In the long-time limit, the second term on the right-hand side converges to zero, and we find the steady-state coherent displacement amplitude ${\beta _{\rm{s}}} = {{{\Omega _b}} \mathord{\left/{\vphantom {{{\Omega _b}} {\left( {{\Delta _{\rm{m}}} - i0.5{\gamma _b}} \right)}}} \right.\kern-\nulldelimiterspace} {\left( {{\Delta _{\rm{m}}} - i0.5{\gamma _b}} \right)}}$. An alternative viewpoint, namely the Gorini-Kossakowski-Sudarshan-Lindblad (GKSL) master equation \cite{ref-60,ref-Lu180401,ref-Li250313731,ref-61} on the evaluation of the steady-state displacement amplitude ${\beta _{\rm{s}}}$ can be found in Appendix \ref{appendix C} \cite{ref-56}.

Within the Born-Markov approximation \cite{ref-62}, the incoming vacuum noise ${{a_{\rm{{in}}}}}$ of the optical cavity field in Eq.~(\ref{eq-8}) and the thermal noise ${{b_{\rm{{in}}}}}$ of the mechanical oscillator in Eq.~(\ref{eq-9}) are fully characterized by the following correlation functions:
\begin{eqnarray}
\langle {a_{{\rm{in}}}}\rangle = 0, \ \langle a_{{\rm{in}}}^\dag \left( t \right){a_{{\rm{in}}}}\left( {t'} \right)\rangle = 0,\nonumber\\ \langle {a_{{\rm{in}}}}\left( t \right)a_{{\rm{in}}}^\dag \left( {t'} \right)\rangle  =  \delta \left( {t - t'} \right),\label{eq-10}\\
\langle {b_{{\rm{in}}}}\rangle  = 0, \ \langle b_{{\rm{in}}}^\dag \left( t \right){b_{{\rm{in}}}}\left( {t'} \right)\rangle  = {{\bar n}_{\rm{m}}}\delta \left( {t - t'} \right)\nonumber,\\
\langle {b_{{\rm{in}}}}\left( t \right)b_{{\rm{in}}}^\dag \left( {t'} \right)\rangle = \left( {{{\bar n}_{\rm{m}}} + 1} \right)\delta \left( {t - t'} \right).\label{eq-11}
\end{eqnarray}
Here, ${{\bar n}_{\rm{m}}}$ is the mean thermal excitation number of the mechanical reservoir under thermal equilibrium. Equations~(\ref{eq-8}) and (\ref{eq-9}) have the form of the nonlinear QLEs since both the light amplitude and the mechanical motion are driven by noise terms that comprise the vacuum noise and the thermal noise entering the system.~Together with Eqs.~(\ref{eq-10}) and (\ref{eq-11}), the nonlinear QLEs describe the evolution of the optical cavity field and the mechanical oscillator, including dissipations $\left( {{\kappa _a},{\gamma _b}} \right)$ and incoming noises $\left( {{a_{\rm{{in}}}},{b_{\rm{{in}}}}} \right)$.

To find the standard form of the nonlinear QLEs, we consider the particular case in which the steady-state displacement amplitude ${{\beta _{\rm{s}}}}$ satisfies the following condition:
\begin{eqnarray}
{\beta _{\rm{s}}} = \beta _{\rm{s}}^* = \left| {{\beta _{\rm{s}}}} \right| = \beta _{\rm{s}}^R > 0. \label{eq-12}
\end{eqnarray}
In other words, the steady-state displacement amplitude ${\beta _{\rm{s}}}$ is real and positive, which is specifically denoted by $\beta _{\rm{s}}^R$. By assuming the laser amplitude ${\Omega _b} = i{\varepsilon _b}\exp \left( {i{\varphi _b}} \right)$, where ${\varepsilon _b}$ denotes the laser power and ${\varphi _b}$ denotes the phase of the laser field coupling to the mechanical mode, we obtain the steady-state solution satisfying Eq.~(\ref{eq-12}) as
\begin{eqnarray}
\beta _{\rm{s}}^R =  - \frac{{2{\varepsilon _b}\cos \left( {{\varphi _b}} \right)}}{{{\gamma _b}}} \ge 0, \quad \tan \left( {{\varphi _b}} \right) = \frac{{2{\Delta _{\rm{m}}}}}{{{\gamma _b}}}\ge 0  \label{eq-13}
\end{eqnarray}
with ${\varphi _b} \in [ {\pi , 1.5\pi } )$. Meeting the above conditions, we arrive at the standard nonlinear QLEs as follows:
\begin{eqnarray}
\dot a \!\!&=& \! \!\!- \left( {i\Delta _{\rm{c}}^0 + \frac{{{\kappa _a}}}{2}} \right)a \!+\! i{g}\left( {{b^\dag } + b} \right)a \!+\! \sqrt {{\kappa _a}} {a_{\rm{in}}} \!-\! i{\Omega _a},\label{eq-14}\\
\dot b \!\!&=& \! \!\!- \left( {i{\Delta _{\rm{m}}} + \frac{{{\gamma _b}}}{2}} \right)b + i{g}{a^\dag }a + \sqrt {{\gamma _b}} {b_{\rm{in}}},\label{eq-15}
\end{eqnarray}
where ${g_{\rm{s}}} = g_{\rm{s}}^* = g = \chi \beta _{\rm{s}}^R$ is the controllable optomechanical coupling strength in the few-photon regime.

\section{Symmetrical optomechanical dynamics at single-photon level}\label{section3}

Before presenting the dynamical analysis, we first emphasize that the construction of a symmetric optomechanical system serves as a framework and provides a baseline for analyzing multimode asymmetric optomechanical systems. Within this symmetric optomechanical framework, we show that the photon and phonon modes exhibit fluctuation dynamics of analogous forms.~We then analyze the optimal reciprocal transport of the input laser field and identify the critical boundary associated with the onset of different coupling regimes. Furthermore, we compare the optical scattering behavior in dissipative equilibrium and nonequilibrium symmetric systems, both with and without non-rotating-wave contributions. Notably, we generalize the scattering-probability expression of Ref.~\cite{ref-81} to a broadly applicable form that is valid not only in the weak-coupling regime but also accurately captures the strong- and ultrastrong-coupling regimes.

\subsection{Real optomechanical coupling}

The nonlinear QLEs.~(\ref{eq-14}) and (\ref{eq-15}) are inherently nonlinear, as they contain products of photon (i.e., optical cavity mode $a$) and phonon (i.e., mechanical mode $b$) operators, ${ab}$ and ${a{b^\dag }}$, as well as a quadratic term in the photon operators, $a^\dagger a$. To proceed, we apply the linearization procedure to Eqs.~(\ref{eq-14}) and (\ref{eq-15}). To this end, we decompose the photon and phonon operators into their classical mean values [i.e., average coherent amplitudes $\langle a \rangle  \approx \alpha$ ($\langle {{a^\dag }} \rangle  \approx {\alpha ^*}$) and $\langle b \rangle  \approx \beta $ ($\langle {{b^\dag }}\rangle  \approx {\beta ^*}$)] and quantum fluctuation operators [i.e., $\delta a$ ($\delta {a^\dag }$) and $\delta b$ ($\delta {b^\dag }$)], namely, $a=\alpha+\delta a$ ($a^\dagger=\alpha^*+\delta a^\dagger$) and $b=\eta+\delta b$ ($b^\dagger=\eta^*+\delta b^\dagger$) \cite{ref-2}. Following this approach, the solution of the complex mean values for the classical steady state satisfies
\begin{eqnarray}
\dot \alpha  &=&  - \left( {i\Delta _c^0 + \frac{{{\kappa _a}}}{2}} \right)\alpha  + i{g}\left( {{\eta ^*} + \eta } \right)\alpha  - i{\Omega _a},\label{eq-16}\\
\dot \eta  &=&  - \left( {i{\Delta _{\rm{m}}} + \frac{{{\gamma _b}}}{2}} \right)\eta  + i{g}{\left| \alpha  \right|^2}.\label{eq-17}
\end{eqnarray}
By solving Eqs.~(\ref{eq-16}) and (\ref{eq-17}), we obtain the solution of the complex mean values $\alpha $ and $\eta $ for the classical steady state as follows:
\begin{eqnarray}
\alpha  &=& \frac{{ - i{\Omega _a}\left[ {{{\left( {\frac{{{\gamma _b}}}{2}} \right)}^2} + {{\left( {{\Delta _{\rm{m}}}} \right)}^2}} \right]}}{{\left( {i\Delta _{\rm{c}}^0 + \frac{{{\kappa _a}}}{2}} \right)\left[ {{{\left( {\frac{{{\gamma _b}}}{2}} \right)}^2} + {{\left( {{\Delta _{\rm{m}}}} \right)}^2}} \right] - 2i{g^2}{\Delta _{\rm{m}}}{{\left| \alpha  \right|}^2}}}, \nonumber \\ \eta  &=& \frac{{ig{{\left| \alpha  \right|}^2}}}{{i{\Delta _{\rm{m}}} + \frac{{{\gamma _b}}}{2}}}.\label{eq-18}
\end{eqnarray}
On the other hand, the quantum fluctuation operators $\delta a$ and $\delta b$ satisfy the following nonlinear equations of motion:
\begin{eqnarray}
\delta \dot a &=&  - \left( {i{\Delta '_{\rm{c}}} + \frac{{{\kappa _a}}}{2}} \right)\delta a + i{g}\alpha \left( {\delta {b^\dag } + \delta b} \right) \nonumber\\&&+ i{g}\delta a\left( {\delta {b^\dag } + \delta b} \right) + \sqrt {{\kappa _a}} \delta {a_{\rm{in}}}, \label{eq-19}\\
\delta \dot b &=&  - \left( {i{\Delta _{\rm{m}}} + \frac{{{\gamma _b}}}{2}} \right)\delta b + i{g}\left(\alpha \delta {a^\dag } + {{\alpha ^*}\delta a} \right) \nonumber\\&&+ i{g}\delta {a^\dag }\delta a + \sqrt {{\gamma _b}} \delta {b_{\rm{in}}},\label{eq-20}
\end{eqnarray}
where the detuning frequency of the optical cavity field is renormalized as ${{\Delta '_{\rm{c}}}} = \Delta _{\rm{c}}^0 - g\left( {\eta  + {\eta ^*}} \right)$.

Since the optical mode is weakly driven by a coherent laser field, while the mechanical mode is strongly coherently driven, as shown in Fig.~\ref{Fig-2}(a), a coherent background generated by dual laser driving and involving a few photons and a finite number of phonons is established (details are provided in Sec.~\ref{section2}). Given this physical scenario, the linearized description is valid provided that the number fluctuations of the photon and phonon modes remain sufficiently small (in the ultrastrong optomechanical coupling regime under few-photon conditions, they differ by about an order of magnitude or more; see Sec.~\ref{section3-B} for a detailed analysis) compared to their corresponding average populations, i.e., $\langle \delta a^\dagger \delta a \rangle \ll \langle a^\dagger a \rangle \approx |\alpha|^2$ and $\langle \delta b^\dagger \delta b \rangle \ll \langle b^\dagger b \rangle \approx |\eta|^2$. Thus, the dominant contributions in Eqs.~(\ref{eq-19}) and (\ref{eq-20}) arise from the first-order linear terms $i{g}\alpha \left( {\delta {b^\dag } + \delta b} \right)$ and $i{g}\left(\alpha \delta {a^\dag } + {{\alpha ^*}\delta a} \right)$. By neglecting the second-order nonlinear terms $ig\,\delta a(\delta b^\dagger+\delta b)$ and $ig\,\delta a^\dagger\delta a$, Eqs.~(\ref{eq-19}) and (\ref{eq-20}) can be linearized, yielding the linearized QLEs for $\delta a$ and $\delta b$:
\begin{eqnarray}
\! \delta \dot a \!\!&=&\!\!\!  - \!\left( {i{\Delta '_{\rm{c}}} \!+\! \frac{{{\kappa _a}}}{2}} \right)\delta a \!+\! i{g}\alpha \left( {\delta {b^\dag } \!+\! \delta b} \right) \!+\!\! \sqrt {{\kappa _a}} \delta {a_{\rm{in}}}, \label{eq-21}\\
\! \delta \dot b \!\!&=&\!\!\!  - \!\left( {i{\Delta _{\rm{m}}} \!+\! \frac{{{\gamma _b}}}{2}} \right)\delta b \!+\! i{g}\left( \alpha \delta {a^\dag } \!+\! {{\alpha ^*}\delta a} \right) \!+\!\! \sqrt {{\gamma _b}} \delta {b_{\rm{in}}}, \label{eq-22}
\end{eqnarray}
which are easy to solve analytically in the frequency domain after Fourier transformation. Furthermore, by observing the form of Eqs.~(\ref{eq-21}) and (\ref{eq-22}), we find that Eq.~(\ref{eq-21}), which describes the dynamical evolution of the photon fluctuation operator, and Eq.~(\ref{eq-22}), which depicts the dynamical behavior of the phonon fluctuation operator, are completely symmetrized when we meet the following two conditions simultaneously in this open quantum system.

The first condition is to set the average value $\alpha $ of the photons in the classical steady state to a real number ${\alpha _R}$. With this restriction, we can determine the real mean value of the classical steady state of the photons ${\alpha _R}$ and ${\eta _{{\alpha _R}}}$ (the value of $\eta $ for a fixed value ${\alpha _R}$) from Eqs.~(\ref{eq-16}) and (\ref{eq-17}) as follows:
\begin{eqnarray}
{\varepsilon _a}\cos \left( {{\varphi _a}} \right) - \frac{{{\kappa _a}}}{2}{\alpha _R} &=&0, \label{eq-23}\\
i{g}{\left( {{\alpha _R}} \right)^2} - \left( {i{\Delta _{\rm{m}}} + \frac{{{\gamma _b}}}{2}} \right){\eta _{{\alpha _R}}} &=& 0, \label{eq-24}\\
{\varepsilon _a}\sin \left( {{\varphi _a}} \right) - {\alpha _R}{\Delta ''_{\rm{c}}} &=& 0, \label{eq-25}
\end{eqnarray}
where we set the laser amplitude ${\Omega _a}$ to $i{\varepsilon _a}\exp \left( {i{\varphi _a}} \right)$ with ${\varepsilon _a}$ denoting the laser power and ${{\varphi _a}}$ the phase of the laser field coupling to the optical mode. We also defined ${\Delta ''_{\rm{c}}} = \Delta _{\rm{c}}^0 - g\left( {{\eta _{{\alpha _R}}} + \eta _{{\alpha _R}}^*} \right)$ as a modified detuning frequency of the optical cavity field. We transform Eqs.~(\ref{eq-23})-(\ref{eq-25}) as follows:
\begin{eqnarray}
{\alpha _R} &=& \frac{{2{\varepsilon _a}\cos \left( {{\varphi _a}} \right)}}{{{\kappa _a}}}, \ \ {\eta _{{\alpha _R}}} = \frac{{2ig{\left( {{\alpha _R}} \right)^2}}}{{2i{\Delta _{\rm{m}}} + {\gamma _b}}},\label{eq-26}\\
\tan \left( {{\varphi _a}} \right) &=& \frac{{2{\Delta ''_{\rm{c}}}}}{{{\kappa _a}}} = \frac{2}{{{\kappa _a}}}\left[ {\Delta _{\rm{c}}^0 - g\left( {{\eta _{{\alpha _R}}} + \eta _{{\alpha _R}}^*} \right)} \right].\label{eq-27}
\end{eqnarray}
Substituting Eq.~(\ref{eq-26}) into Eq.~(\ref{eq-27}), we obtain a nonlinear equation for ${{\varphi _a}}$ in the form
\begin{eqnarray}
\!\!\tan \left( {{\varphi _a}} \right) \!=\!\! \frac{2}{{{\kappa _a}}}\!\!\left\{ {\Delta _{\rm{c}}^0 \!\!-\!\! \frac{{8{\Delta _{\rm{m}}}}}{{4{{\left( {{\Delta _{\rm{m}}}} \right)}^2} \!\!+\!\! {{\left( {{\gamma _b}} \right)}^2}}}{{\left[ {\frac{{2g{\varepsilon _a}\cos \left( {{\varphi _a}} \right)}}{{{\kappa _a}}}} \right]}^2}} \right\}\!.\label{eq-28}
\end{eqnarray}
Since ${\alpha _R} > 0$, we find the solution of Eq.~(\ref{eq-28}) in the range: ${\varphi _a} \in \left[ {0,0.5\pi } \right) \cup \left( {1.5\pi ,2\pi } \right]$. We note that, when $\alpha$ is taken as a complex mean value satisfying the classical steady-state solution, its phase may break the symmetry of the dynamics and induce nonreciprocal transmission in general. Although this mechanism is not operative in a two-mode case (since the physical phase angle can be eliminated by a suitable transformation of the reference frame), it can become relevant in more general multimode configurations \cite{ref-65,ref-Yang2505.10255,ref-Yi2503.23169,ref-Luan2503.18647}.

Now, substituting the expressions~(\ref{eq-26}) into Eqs.~(\ref{eq-21}) and (\ref{eq-22}), we can reduce them to the following forms:
\begin{eqnarray}
\delta \dot a \!\!&=&\!\!\!  - \!\left( {i{{\Delta ''_{\rm{c}}}} + \frac{{{\kappa _a}}}{2}} \right)\delta a \!+\! i{G_R}\left( {\delta {b^\dag } + \delta b} \right) \!+\!\! \sqrt {{\kappa _a}} \delta {a_{\rm{in}}}, \label{eq-29}\\
\delta \dot b \!\!&=&\!\!\!  - \!\left( {i{\Delta _{\rm{m}}} + \frac{{{\gamma _b}}}{2}} \right)\delta b \!+\! i{G_R}\left( {\delta {a^\dag } + \delta a} \right) \!+\!\! \sqrt {{\gamma _b}} \delta {b_{\rm{in}}},\label{eq-30}
\end{eqnarray}
where ${G_R} = {g}{\alpha _R}$ is an effective optomechanical coupling in the linearized regime. It is enhanced compared to ${g}$ by the real amplitude ${\alpha _R}$ of the photon field.

Equations~(\ref{eq-29}) and (\ref{eq-30}) show the formal equivalence of the quantum fluctuation dynamics of photons and phonons. The corresponding effective Hamiltonian of the system under the Heisenberg picture (time-dependent operator) is given by ${H_{{\rm{eff}}}} = \hbar {\Delta ''_{\rm{c}}}{a^\dag }a + \hbar {\Delta _{\rm{m}}}{b^\dag }b - \hbar {G_R}\left( {{a^\dag } + a} \right)\left( {{b^\dag } + b} \right)$. Hereafter, we refer to ${H_{{\rm{eff}}}}$ as the effective Hamiltonian of the symmetric optomechanical system, in which the interaction form describes a general linear coupling between two bosonic modes in quantum optics.~Moreover, we know that the Hamiltonian of the Jaynes-Cummings model describing the atom-field interaction has a similar interaction form $\hbar g({a^\dag } + a)({\sigma _ + } + {\sigma _ - })$, where the operator ${{\sigma _ + }}$ takes an atom in the lower state into the upper state whereas ${{\sigma _ - }}$ has the opposite effect. Thus, the effective symmetric Hamiltonian ${H_{{\rm{eff}}}}$ provides a general formalism for linear light-matter interactions in quantum optics, and the resulting conclusions are of broad applicability.

Another condition for achieving the equivalence of the quantum fluctuation dynamics of the photon and phonon modes is to set the corresponding parameters in Eqs.~(\ref{eq-29}) and (\ref{eq-30}) equal to each other. To be specific, we equalize the modified detuning frequency ${{\Delta ''_c}}$ of the optical cavity field to the detuning of the mechanical resonance frequency ${{\Delta _{\rm{m}}}}$ by adjusting the two driving lasers ${\Omega _a}$ and ${\Omega _b}$. We also equalize the decay ${{\kappa _a}}$ of the optical cavity field to the damping ${{\gamma _b}}$ of the mechanical oscillator by modulating the corresponding system-reservoir interaction ${g_{a{\rm{k}}}}$ and ${g_{b{\rm{k}}}}$. To implement this scenario, we carefully analyze and compare representative works in the theoretical proposal and experimental results concerning the optomechanical systems with the enhanced CK interaction; see Table~\ref{tab-1}. We see that circuit quantum electrodynamic systems provide a practical platform for realizing the above scenario, where the parameter conditions ${\Delta ''_{\rm{c}}} = {\Delta _{\rm{m}}}$ and ${\kappa _a} = {\gamma _b}$ are experimentally achievable. In circuit QED, the cavity decay rate ${\kappa _a}$ can be tuned via the external coupling between the resonator and transmission lines, while the effective mechanical damping rate ${\gamma _b}$ with parametric mechanical driving can be independently engineered through dissipative coupling or reservoir engineering, allowing the condition ${\kappa _a} = {\gamma _b}$ to be experimentally achieved \cite{ref-exp-1,ref-exp-2}. Additionally, the condition ${\Delta ''_{\rm{c}}} = {\Delta _{\rm{m}}}$ can be achieved experimentally by adjusting the driving fields of the optical and mechanical modes so that the resonance parameters are matched.

\begin{table*}[htbp]
	\centering
    \setlength{\abovecaptionskip}{0pt}
    \setlength{\belowcaptionskip}{8pt}
    \renewcommand{\arraystretch}{1}
    \setlength\tabcolsep{3.4mm}
	\caption{ Representative works in theoretical proposals and experimental results for various optomechanical systems with the CK interaction. Note that the parameters listed in the table represent typical ranges extracted from different experiments and are compatible. Typical parameters of the standard cavity optomechanical system satisfy ${\Delta ''_{\rm{c}}} = {\Delta _{\rm{c}}}$. The first line shows the maximum parameter values achieved by the system (\ref{eq-5})-(\ref{eq-6}) simulated by the QED circuit after laser adjustment.}
	\label{tab-1}
    \begin{tabular}{ccccc ccccc}
		\hline\hline\noalign{\smallskip}
         References & Type $\left( {\rm{{Hz}}} \right)$ & ${{{{\Delta ''_{\rm{c}}}}} \mathord{\left/{\vphantom {{{{\Delta ''}_{\rm{c}}}} {2\pi }}} \right.\kern-\nulldelimiterspace} {2\pi }}$ & ${{{\Delta _{\rm{m}}}} \mathord{\left/ {\vphantom {{{\Delta _{\rm{m}}}} {2\pi }}} \right.\kern-\nulldelimiterspace} {2\pi }}$ & ${{{\kappa _a}} \mathord{\left/{\vphantom {{{\kappa _a}} {2\pi }}} \right.\kern-\nulldelimiterspace} {2\pi }}$ & ${{{\gamma _b}} \mathord{\left/{\vphantom {{{\gamma _b}} {2\pi }}} \right.\kern-\nulldelimiterspace} {2\pi }}$ & ${\chi  \mathord{\left/ {\vphantom {\chi  {2\pi }}} \right. \kern-\nulldelimiterspace} {2\pi }}$\\
         \hline\noalign{\smallskip}
          \cite{ref-49,ref-56} & Circuit-QED$  $ & $5.2 \times {10^9}$ & $5.2 \times {10^9}$ & $3.3 \times {10^7}$ \cite{ref-52} & $3.3 \times {10^7}$ \cite{ref-52} & $2.5  \times {10^6}$ \cite{ref-50}\\
          \cite{ref-66,ref-67} & Cavity-QED$   $ & $4.8 \times {10^9}$ & $7.2 \times {10^7}$ & $ 4.2 \times {10^7}$ & $2.6 \times {10^6}$ & ${10^4} \sim {10^5}$\\
          \cite{ref-68,ref-69} & Cavity-optom$   $& $1.1 \times {10^7}$ & $7.8 \times {10^7}$ & $7.1 \times {10^6}$ & $3.4 \times {10^3}$ & $ \approx 0$\\
          \cite{ref-70,ref-71} & Quadratic-optom$  $& $2.8 \times {10^{14}}$ & $1.3 \times {10^6}$ & $5.9 \times {10^5}$ & $1.2 \times {10^{ - 1}}$ & $3.95 \times {10^{ - 2}}$  \\
         \hline\hline\noalign{\smallskip}
         \end{tabular}
\end{table*}

When the two conditions mentioned above are satisfied, the quantum fluctuation dynamics of photons and phonons are completely symmetrized. Equations~(\ref{eq-29}) and (\ref{eq-30}) now read
\begin{eqnarray}
\delta \dot a \!\!&=&\!\!  - \left( {i\Delta  + \frac{\xi }{2}} \right)\delta a + i{G_R}\left( {\delta {b^\dag } + \delta b} \right) \!+\! \sqrt \xi  \delta {a_{\rm{in}}}, \label{eq-31}\\
\delta \dot b \!\!&=&\!\!  - \left( {i\Delta  + \frac{\xi }{2}} \right)\delta b + i{G_R}\left( {\delta {a^\dag } + \delta a} \right) \!+\! \sqrt \xi  \delta {b_{\rm{in}}},\label{eq-32}
\end{eqnarray}
where $\Delta  = {\Delta ''_{\rm{c}}} = {\Delta _{\rm{m}}}$ and $\xi  = {\kappa _a} = {\gamma _b}$. Equations~(\ref{eq-31}) and (\ref{eq-32}) describe the quantum fluctuation dynamics of the completely symmetric optomechanics in open quantum systems. The symmetrization of the quantum fluctuation dynamics of photons and phonons reveals a more profound connection under quantum mechanics; namely, both are bosons.

\subsection{Implementation of optomechanical ultrastrong coupling in the coherent-state representation}\label{section3-B}

\begin{figure}[t]
\centering
\includegraphics[angle=0,width=0.44\textwidth]{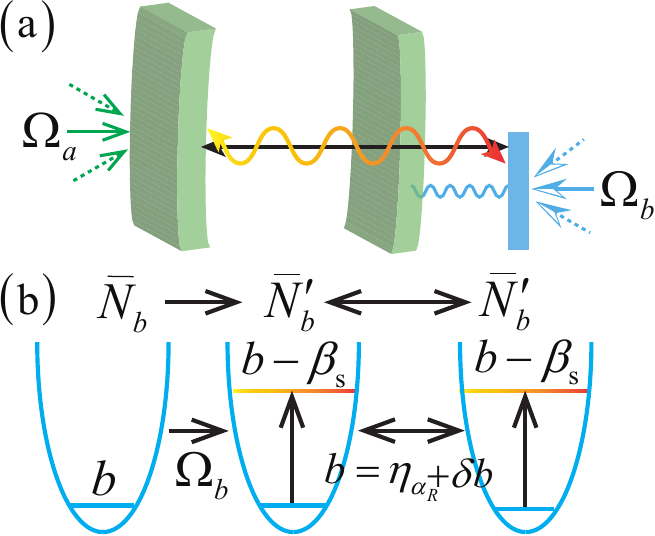}
\caption{$\left( {\rm{a}} \right)$ Schematic diagram of a symmetric optomechanical system with dual laser drives. $\left( {\rm{b}} \right)$ The number of excited phonons in the system after the coherent displacement transformation at the steady state and the standard mean-field operation.} \label{Fig-2}
\end{figure}

In the symmetric optomechanical Eqs.~(\ref{eq-31}) and (\ref{eq-32}), as shown in Fig.~\ref{Fig-2}, we adopt the form of the dual coherent laser driving in order to achieve an optomechanical ultrastrong coupling in the coherent-state representation (a detailed physical interpretation is presented in Appendix~\ref{appendix C}). On the one hand, a lower-power monochromatic laser ${\Omega _a}$ drives the optical cavity $a$ to excite a small number of photons ${\bar N_a}$ in the optical cavity. On the other hand, a high-power monochromatic laser ${\Omega _b}$ drives the mechanical oscillator $b$ so that steady-state displacement amplitude ${\beta _{\rm{s}}}$ may be large enough to achieve ultrastrong optomechanical coupling. In this work, we access the ultrastrong optomechanical coupling regime with a small number of photons and a finite number of phonons by dual coherent laser driving. For practical implementation, two requirements need to be appropriately considered. First, the power of the laser field driving the optical cavity should be properly chosen such that the intracavity photon number remains small, ensuring operation in the few-photon regime. The second requirement is to determine the relationship between the laser power driving the mechanical oscillator and its steady-state displacement amplitude, in order to achieve ultrastrong optomechanical coupling.

First, we evaluate the relationship between the number of photons in the system and the low-power driving laser. We know from Eq.~(\ref{eq-26}) that the average photon number acting on the symmetrical optomechanical system is given by
\begin{eqnarray}
{{\bar N}_a} =  {\alpha _R}^2 = 2{\left( {\frac{{{\varepsilon _a}}}{{{\kappa _a}}}} \right)^2}\left[ {\cos \left( {2{\varphi _a}} \right) + 1} \right], \label{eq-33}
\end{eqnarray}
in which ${\varepsilon _a}$ is related to the driving laser power ${P_a}$ by ${\varepsilon _a} = \left| {{\Omega _a}} \right| = \sqrt {{{2{P_a}{\kappa _a}} \mathord{\left/{\vphantom {{2{P_a}{\kappa _a}} {\hbar {\omega _{{L_a}}}}}} \right.\kern-\nulldelimiterspace} {\hbar {\omega _{{L_a}}}}}}$. Equation~(\ref{eq-33}) can therefore be rewritten as
\begin{eqnarray}
0 \le {{\bar N}_a} = \frac{{4{P_a}\left[ {\cos \left( {2{\varphi _a}} \right) + 1} \right]}}{{\hbar {\omega _{{L_a}}}{\kappa _a}}} \le \frac{{8{P_a}}}{{\hbar {\omega _{{L_a}}}{\kappa _a}}}. \label{eq-34}
\end{eqnarray}

\begin{figure}[t]
\centering
\includegraphics[angle=0,width=0.34\textwidth]{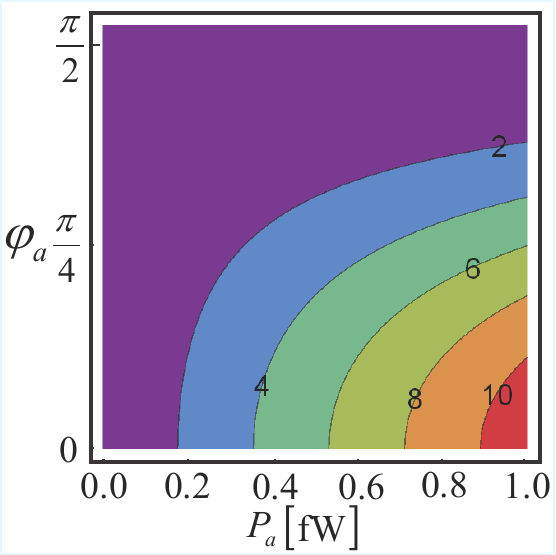}
\caption{Average photon number $\bar N_a$ in the optical cavity as a function of the coherent driving laser power $P_a \in [0,\,1\,\mathrm{fW}]$ and the phase $\varphi_a \in [0,\,0.5\pi)$ of the optical drive $\Omega_a$. The parameters are taken from the circuit-QED setup listed in Table~\ref{tab-1}: $\Delta_{\rm m} = \omega_{L_a} = \omega_{L_b} = 1.04\pi \times 10^{10}\,\mathrm{Hz}$, $\kappa_a = \gamma_b = 6.6\pi \times 10^{7}\,\mathrm{Hz}$, $\chi = 5\pi \times 10^{6}\,\mathrm{Hz}$, and $\hbar = 1.05 \times 10^{-34}\,\mathrm{J\cdot s}$.} \label{Fig-3}
\end{figure}

For simplicity, we assume that the driving laser frequency $\omega_{L_a}$ of the optical cavity field is resonant with the modified cavity detuning $\Delta''_{\rm c}$. In Fig.~\ref{Fig-3}, we show the dependence of the average intracavity photon number $\bar N_a$ on the driving laser power $P_a$ and the phase $\varphi_a$ of the optical drive $\Omega_a$. As shown in Fig.~\ref{Fig-3}, by tuning the phase $\varphi_a$ within the range $[0,\,0.5\pi]$ and choosing the driving power $P_a$ at the fW level, the optical cavity can be operated in the few-photon regime, where the intracavity photon number remains small. The average intracavity photon number $\bar N_a$ is proportional to the driving power $P_a$, except at phases $\varphi_a = \pi/2 + n\pi$ ($n\in\mathbb{Z}$), where the optical drive does not couple into the cavity.~For instance, at $\varphi_a = 0$, one has ${\bar N_a} = {{8{P_a}} \mathord{\left/
 {\vphantom {{8{P_a}} {\hbar {\omega _{{L_a}}}{\kappa _a}}}} \right.
 \kern-\nulldelimiterspace} {\hbar {\omega _{{L_a}}}{\kappa _a}}}$. Below, the intracavity photon number is chosen as ${\bar N_a} \sim 10$ (${\alpha _R} \approx 3.16$).

\begin{table*}[htbp]
	\centering
    \setlength{\abovecaptionskip}{0pt}
    \setlength{\belowcaptionskip}{8pt}
    \renewcommand{\arraystretch}{1}
    \setlength\tabcolsep{2mm}
	\caption{Regimes of optomechanical interaction \cite{ref-76} and corresponding ${\beta _{\rm{s}}^R}$, ${P_b}\left( {{\varphi _b} = \pi } \right)$, and ${{\bar N}_b}$ for ${{\bar N}_a} \sim 10$.}
	\label{tab-2}
	\begin{tabular}{ccccc cccccc}
	    \hline\hline\noalign{\smallskip}	
		 & ${G_R}$ & ${g_0}$ & $0.5{\kappa _a}$ & $0.1{\Delta _{\rm{m}}}$ & ${\Delta _{\rm{m}}}$ & $ > {\Delta _{\rm{m}}}$ & $\beta _{\rm{s}}^R$ & ${{P_b}}$ & ${{\bar N}_b}$ \\
        \noalign{\smallskip}\hline\noalign{\smallskip}
         & Weak  & $ \bullet $ & $ \circ $ &  &  &  & $\left[ {0,2.1} \right)$ & $\left[ {0,0.39{\rm{fW}}} \right)$ & $\left( {0,4} \right]$\\
		 & Strong  &  & $ \bullet $ & $ \circ $ &  &  & $\left[ {2.1,65.8} \right)$ & $\left[ {0.39{\rm{fW}},0.385{\rm{pW}}} \right)$ & $\left[ {5,4.32 \times {10^3}} \right)$  \\
         & Ultrastrong &  &  & $ \bullet $ &  $ \bullet $ &  & $\left[ {65.8,657.8} \right)$ & $\left[ {0.385{\rm{pW}},38.5{\rm{pW}}} \right)$ & $\left[ {4330,4.33 \times {{10}^5}} \right)$ \\
         & Deep strong  &  &   &  &  & $ * $ &  $ > 658$  & $ > 38.5{\rm{pW}}$ & $ > 4.33 \times {10^5}$\\
        \hline\hline\noalign{\smallskip}
   \end{tabular}
\end{table*}

Furthermore, we determine the required laser power $P_b$ to achieve optomechanical ultrastrong coupling characterized by the effective coupling strength $G_R$. Experimentally, optomechanical interactions are commonly classified into four coupling regimes, namely the weak \cite{ref-72}, strong \cite{ref-73}, ultrastrong \cite{ref-74}, and deep strong \cite{ref-75} regimes. Accordingly, we calculate the range of the driving power $P_b$ corresponding to each of these four regimes. Here, the effective optomechanical coupling has the form ${G_R} = g{\alpha _R}$ with $g = \chi {\beta _{\rm{s}}^R}$. From Eq.~(\ref{eq-13}), we know that ${\beta _{\rm{s}}^R}$ is proportional to the laser power by ${\varepsilon _b} = \left| {{\Omega _b}} \right| = \sqrt {{{2{P_b}{\gamma _b}} \mathord{\left/{\vphantom {{2{P_b}{\gamma _b}} {\hbar {\omega _{{L_b}}}}}} \right.\kern-\nulldelimiterspace} {\hbar {\omega _{{L_b}}}}}} $. The expression~(\ref{eq-13}) is therefore written as
\begin{eqnarray}
{\beta _{\rm{s}}^R} =  - \frac{{2{\varepsilon _b}\cos \left( {{\varphi _b}} \right)}}{{{\gamma _b}}} =  - \frac{{2\cos \left( {{\varphi _b}} \right)}}{{{\gamma _b}}}\sqrt {\frac{{2{P_b}{\gamma _b}}}{{\hbar {\omega _{{L_b}}}}}}  \label{eq-35}
\end{eqnarray}
with ${\varphi _{\rm{b}}} \in [ {\pi ,1.5\pi } )$. Similarly, we assume that the driving laser frequency ${{\omega _{{L_b}}}}$ of the mechanical mode is resonant with the detuning of mechanical resonance frequency ${\Delta _{\rm{m}}}$. Let us use the data of the circuit-QED in Table~\ref{tab-1}: ${\kappa _a}={\gamma _b} = 6.6\pi  \times {10^7}{\rm{Hz}},{\omega _{{L_b}}} = {\Delta _{\rm{m}}} = 1.04\pi  \times {10^{10}}{\rm{Hz}},\chi  = 5\pi  \times {10^6}{\rm{Hz}}$, and $\hbar {\rm{ = 1}}{\rm{.05}} \times {\rm{1}}{{\rm{0}}^{ - 34}}J \cdot s$. As shown in Table~\ref{tab-2}, for $\bar N_a \sim 10$ (${\alpha _R} \approx 3.16$) and ${\varphi _b} = \pi $, achieving the ultrastrong optomechanical coupling regime ${G_R} \sim 0.1{\Delta _{\rm{m}}}$ requires a laser power $P_b$ exceeding $384.97\,\mathrm{fW}$, corresponding to $\beta_{\rm s}^R \approx 65.8$. The analysis shows that achieving ultrastrong coupling in the few-photon regime requires weak optical driving and strong mechanical driving.

Finally, let us find the number of phonons ${{\bar N}_b}$ for each regime of ${P_b}$. From Eqs.~(\ref{eq-26}) and~(\ref{eq-35}) we know that the average phonon number working on the symmetric optomechanics is given by
\begin{eqnarray}
{{\bar N}_b} &=& {\left| {\beta _{\rm{s}}^R} \right|^2} + {\left| {{\eta _{{\alpha _R}}}} \right|^2} = {\left| {\beta _{\rm{s}}^R} \right|^2} + \frac{{{{\left( {{G_R}} \right)}^2}}}{{{{\left( {\frac{{{\gamma _b}}}{2}} \right)}^2} + {{\left( {{\Delta _{\rm{m}}}} \right)}^2}}}\nonumber\\
&=& \frac{{8{P_b}{{\left[ {\cos \left( {{\varphi _b}} \right)} \right]}^2}}}{{\hbar {\omega _{{L_b}}}{\gamma _b}}}\left[ {1 + \frac{{{\left| {{\alpha _R}} \right|^2}{\chi ^2}}}{{{{\left( {\frac{{{\gamma _b}}}{2}} \right)}^2} + {{\left( {{\Delta _{\rm{m}}}} \right)}^2}}}} \right].\label{eq-36}
\end{eqnarray}

From Table~\ref{tab-2} and Eq.~(\ref{eq-36}), we find the number of phonons in the four regimes: in the weak coupling regime ${{\bar N}_b} \!\in \left( {0,4} \right]$; in the strong coupling regime ${{\bar N}_b}\! \in \left[ {5,4329 }\right)$; in the ultrastrong coupling regime ${{\bar N}_b}\! \in \left[ {4330,4.33 \times {{10}^5}} \right)$; in the deep coupling regime ${{\bar N}_b}$ more than ${4.33 \times {{10}^5}}$. These results demonstrate that, within the few-photon regime, the ultrastrong optomechanical coupling can be achieved through interactions between a few photons and a finite number of thermally excited phonons.

To summarize, ultrastrong optomechanical coupling in the few-photon regime can be achieved in a circuit-QED platform with a weak coherent optical drive at the fW level ($P_a \sim \mathrm{fW}$) and a comparatively stronger coherent mechanical drive at the pW level ($P_b \sim \mathrm{pW}$), for $\varphi_a = \pi$ and $\varphi_b = \pi$. This operating condition is fully consistent with the working assumption of strong coherent mechanical driving and weak coherent optical driving ($P_b \gg P_a$), which enables ultrastrong optomechanical coupling.

\section{Input-Output theory under strong and ultrastrong coupling regimes}\label{section4}
\begin{figure}[h]
\centering
\includegraphics[angle=0,width=0.45\textwidth]{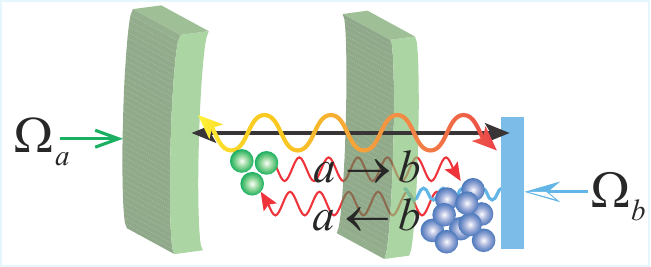}
\caption{Schematic diagram of a symmetric optomechanical system with dual coherent laser drives operating in the few-photon regime with a finite number of phonons. The wavy line with $a \to b$ represents the transmission process of the photon signal from the optical cavity field to the mechanical oscillator, while the one with $b \to a$ describes the opposite process of the phonon signal.} \label{Fig-4}
\end{figure}
In the preceding section, we obtain the Heisenberg-Langevin equation for a symmetric optomechanical system with a controllable optomechanical coupling operating in the few-photon regime. In this section, considering the input laser field shown in Fig.~\ref{Fig-4}, we further derive the input-output formula of the incident laser field and study its scattering characteristics in the open quantum system. First, we numerically verify the conditions under which the rotating-wave approximation (RWA) holds in this system. Secondly, we give the conditions of the optimal reciprocal transmission of the incoming laser field in the system. We then compare the scattering behavior of the incident laser field in the dissipative balanced and unbalanced symmetric optomechanics before and after the RWA.

\subsection{Applicability Conditions for Numerical Evaluation under RWA}

In this part, we give the scattering matrix theory and evaluate the applicability of RWA by numerical calculation.

In the case without RWA, for convenience, we concisely express the linearized QLEs.~(\ref{eq-29})-(\ref{eq-30}) as
\begin{eqnarray}
\dot V =  - {\rm K}V + {\mu }{V_{\rm{{in}}}}, \label{eq-38}
\end{eqnarray}
where the component of each matrix is as follows: the quantum fluctuation operator is written as $V = {( {\delta a,\delta b,\delta {a^\dag },\delta {b^\dag }} )^T}$; the noise field operator is denoted by ${V_{{\rm{in}}}} = {(\delta {a_{{\rm{in}}}},\delta {b_{{\rm{in}}}},\delta a_{{\rm{in}}}^\dag ,\delta b_{{\rm{in}}}^\dag )^T}$; the damping operators read ${\mu } = {\rm{diag}}( {\sqrt {{\kappa _a}} ,\sqrt {{\gamma _b}} \sqrt {{\kappa _a}} ,\sqrt {{\gamma _b}} })$; the coefficient matrix ${\rm K}$ in terms of system parameters takes the form
\begin{eqnarray}
\!\!{\rm K}{\ \rm{=}}\left[ {\begin{array}{*{20}{c}}
{\!i{\Delta ''_c} + \frac{{{\kappa _a}}}{2}}&\!{ - i{G_R}}&\!0&\!{ - i{G_R}}\\
{\! - i{G_R}}&\!{i{\Delta _{\rm{m}}} + \frac{{{\gamma _b}}}{2}}&\!{ - i{G_R}}&\!0\\
\!0&\!{i{G_R}}&\!{ - i{\Delta ''_c} + \frac{{{\kappa _a}}}{2}}&\!{i{G_R}}\\
\!{i{G_R}}&\!0&\!{i{G_R}}&\!{ - i{\Delta _{\rm{m}}} + \frac{{{\gamma _b}}}{2}}
\end{array}} \right]\!\!\!. \label{eq-39}
\end{eqnarray}
The necessary and sufficient condition for the stability of the dynamical system it describes is that the real parts of all the eigenvalues of the coefficient matrix ${\rm K}$ are positive. Physically, this condition implies that any small deviation from the steady state decays exponentially in time, indicating that all eigenmodes of the linearized dynamics are stable and that the system can relax back to the steady state \cite{ref-add-RH}. Mathematically, this condition is equivalent given by the Routh-Hurwitz stability criterion \cite{ref-77}. Specifically, for the eigenvalue equation det$|{\rm{K}} - \lambda {{\rm{I}}_4}| = 0$, where ${{\rm{I}}_4}$ denotes the four-dimensional identity matrix, which can be reduced to the fourth-order equation ${C_0}{\lambda ^4} + {C_1}{\lambda ^3} + {C_2}{\lambda ^2} + {C_3}\lambda  + {C_4}{\rm{ = }}0$, the Routh–Hurwitz criterion indicates that the system is stable only if all of the following conditions are satisfied: ${C_1} > 0,{C_1}{C_2} - {C_0}{C_3} > 0,\left( {{C_1}{C_2} - {C_0}{C_3}} \right){C_3} - C_1^2{C_4} > 0,{C_4} > 0$ \cite{ref-77}. In other words, if any of the sub-conditions (such as ${C_4} < 0$) above is is violated, the system enters an unstable regime. Physically, violation of any of these conditions leads to the exponential amplification of the corresponding fluctuation effects, resulting in the divergence of their amplitudes or the emergence of parametric instability regions. In our work, the parameters are carefully chosen to ensure that the system satisfies the stability conditions. However, we point out that unstable regions exist in parameter space \cite{ref-77}. In detail, in our work, det$|{\rm{K}} - \lambda {{\rm{I}}_4}| = 0$ can be explicitly written as
\begin{widetext}
\begin{eqnarray}
   \left[ {{{\left( {\frac{{{\kappa _a}}}{2} - \lambda } \right)}^2} + {{\left( {{\Delta ''_{\rm{c}}}} \right)}^2}} \right]\left[ {{{\left( {\frac{{{\gamma _b}}}{2} - \lambda } \right)}^2} + {{\left( {{\Delta _{\rm{m}}}} \right)}^2}} \right]  - 4{\Delta _{\rm{m}}}{\Delta ''_{\rm{c}}}{\left( {{G_R}} \right)^2} = 0. \label{eq-RH-add-1}
\end{eqnarray}
By expanding Eq.~\ref{eq-RH-add-1}, we obtain ${C_0}{\lambda ^4} + {C_1}{\lambda ^3} + {C_2}{\lambda ^2} + {C_3}\lambda  + {C_4} = 0$, where
\begin{eqnarray}
{C_0} &=& 1,{C_1} =  - \left( {{\kappa _a} + {\gamma _b}} \right),{C_2} = {\left( {\frac{{{\kappa _a}}}{2}} \right)^2} + {\left( {\frac{{{\gamma _b}}}{2}} \right)^2} + {\kappa _a}{\gamma _b} + {\left( {{\Delta _{\rm{m}}}} \right)^2} + {\left( {{\Delta ''_{\rm{c}}}} \right)^2},\nonumber\\
{C_3} &=&  - \left\{ {\left[ {\frac{1}{4}{{\left( {{\kappa _a}} \right)}^2} + {{\left( {{\Delta ''_{\rm{c}}}} \right)}^2}} \right]{\gamma _b} + \left[ {\frac{1}{4}{{\left( {{\gamma _b}} \right)}^2} + {{\left( {{\Delta _{\rm{m}}}} \right)}^2}} \right]{\kappa _a}} \right\},\label{eq-RH-add-2}\\
{C_4} &=& \frac{1}{4}{\left( {{\kappa _a}} \right)^2}{\left( {{\Delta _{\rm{m}}}} \right)^2} + \frac{1}{4}{\left( {{\gamma _b}} \right)^2}{\left( {{\Delta ''_{\rm{c}}}} \right)^2} + \frac{1}{{16}}{\left( {{\kappa _a}} \right)^2}{\left( {{\gamma _b}} \right)^2} + {\left( {{\Delta ''_{\rm{c}}}} \right)^2}{\left( {{\Delta _{\rm{m}}}} \right)^2} - 4{\Delta _{\rm{m}}}{\Delta ''_{\rm{c}}}{\left( {{G_R}} \right)^2}.\nonumber
\end{eqnarray}
\end{widetext}
It is straightforward to verify that, in the ultrastrong-coupling regime, for parameters chosen as ${G_R} > 0.5{\Delta _{\rm{m}}} \approx 0.5{{\Delta ''}_{\rm{c}}} \gg {\rm{Max}}\left( {{\kappa _a},{\gamma _b}} \right)$, we obtain ${C_4} < 0$, thereby showing an unstable region in parameter space. In the following, our numerical calculations ensure that the stability conditions are satisfied for the parameters used.

By introducing the Fourier transform for an operator $\Lambda \left( t \right)$ as
\begin{eqnarray}
\Lambda \left( \omega  \right) = \mathscr{F}\left[ {\Lambda \left( t \right)} \right] = \frac{1}{{\sqrt {2\pi } }}\int_{ - \infty }^{ + \infty } {\Lambda \left( t \right)} {e^{i\omega t}}dt \label{eq-40}
\end{eqnarray}
and using its derivative property $\mathscr{F}[ {\dot \Lambda  \left( t \right)} ] =  - i\omega \mathscr{F}\left[ {\Lambda \left( t \right)} \right]$, we find the solution to the linearized QLEs.~(\ref{eq-29})-(\ref{eq-30}) in the frequency domain in form
\begin{eqnarray}
V\left( \omega  \right) = {\left[ {{\rm{K}} - i\omega {{\rm{I}}_4}} \right]^{ - 1}}\mu {V_{{\rm{in}}}}\left( \omega  \right), \label{eq-41}
\end{eqnarray}

Under boundary conditions, the relation among the input, internal, and output fields can be given by the input-output theory \cite{ref-78}. Substituting Eq.~(\ref{eq-41}) into the input-output relation ${V_{{\rm{out}}}}\left( \omega  \right) + {V_{{\rm{in}}}}\left( \omega  \right) = {\mu}V\left( \omega  \right)$, we obtain
\begin{eqnarray}
{V_{{\rm{out}}}}\left( \omega  \right) = {\rm{O}}\left( \omega  \right){V_{{\rm{in}}}}\left( \omega  \right),  \label{eq-42}
\end{eqnarray}
in which the output field matrix ${V_{{\rm{out}}}}\left( \omega  \right)$ is the Fourier transform of ${V_{{\rm{out}}}} = {(\delta {a_{{\rm{out}}}},\delta {b_{{\rm{out}}}},\delta a_{{\rm{out}}}^\dag ,\delta b_{{\rm{out}}}^\dag )^T}$. The scattering matrix for the symmetric optomechanical system can be written as
\begin{eqnarray}
{\rm{O}}\left( \omega  \right) = {\mu ^T}{\left[ {{\rm{K}} - i\omega {{\rm{I}}_4}} \right]^{ - 1}}\mu  - {{\rm{I}}_4}.\label{eq-43}
\end{eqnarray}
The spectrum of the output field is defined by
\begin{eqnarray}
{S_{{\rm{out}}}}\left( \omega  \right) = \int {d\omega '} \langle V_{{\rm{out}}}^\dag \left( {\omega '} \right){V_{{\rm{out}}}}\left( \omega  \right)\rangle . \label{eq-44}
\end{eqnarray}
By substituting Eq.~(\ref{eq-42}) into Eq.~(\ref{eq-44}), we have
\begin{eqnarray}
{S_{{\rm{out}}}}\left( \omega  \right) = {\rm{P}}\left( \omega  \right){S_{{\rm{in}}}}\left( \omega  \right) + \Theta_{\rm{vac}} \left( \omega  \right),\label{eq-45}
\end{eqnarray}
where the term ${\rm P}\left( \omega  \right)$ denotes the scattering probability that corresponds to the contribution arising from the input laser field. The term $\Theta_{\rm{vac}} \left( \omega  \right)$ is a contribution of the input vacuum field to the output spectrum (see Ref.~\cite{ref-79} for the explicit definition of the term), which is an effect of the anti-RWA terms \cite{ref-80}. Let us modify the results of Ref.~\cite{ref-81}. The modified scattering probability has broad applicability, which is not only applicable to weak coupling but also suitable to strong coupling and even ultrastrong coupling [see Appendix \ref{appendix D} for details]. The functions ${\rm P}\left( \omega  \right)$ and $\Theta_{\rm{vac}} \left( \omega  \right)$ have the form
\begin{eqnarray}
a \to b: \ {\rm P}_a^b\left( \omega  \right) &=& {\left| {{\rm O}_a^b\left( \omega  \right)} \right|^2}; \nonumber\\ \Theta _{{\rm{vac}}}^a\left( \omega  \right) &=& |{\rm{O}}_a^{{a^\dag }}\left( \omega  \right){|^2} + |{\rm{O}}_a^{{b^\dag }}\left( \omega  \right){|^2},\label{eq-46}\\
a \leftarrow b: \ {\rm P}_b^a\left( \omega  \right) &=& {\left| {{\rm O}_b^a\left( \omega  \right)} \right|^2}; \nonumber\\ \Theta _{{\rm{vac}}}^b\left( \omega  \right) &=& |{\rm{O}}_b^{{a^\dag }}\left( \omega  \right){|^2} + |{\rm{O}}_b^{{b^\dag }}\left( \omega  \right){|^2},\label{eq-47}
\end{eqnarray}
where $a\!\! \to \!\!b$ represents the transmission of the laser signal from the optical mode $a$ to the mechanical mode $b$, while $a\!\leftarrow\!b$ denotes effects in the reverse process, and ${\rm O}_a^b\left( \omega  \right)$ represents the corresponding element at the first $\delta a$ row and the second $\delta b$ column of the scattering matrix ${\rm O}\left( \omega  \right)$ given in Eq.~(\ref{eq-43}).

To calculate the scattering behavior of the incoming laser field in the symmetric optomechanics, we use the scattering matrix theory. For simplicity, taking the parameter values for the QED circuit in Table~\ref{tab-1}, we have ${\Delta ''_{\rm{c}}} = {\Delta _{\rm{m}}} \approx 158{\kappa _a} = 158{\gamma _b}$. These parameters satisfy the validity conditions of the rotating-wave approximation (RWA), that is, both the red detuned regime ${\Delta ''_c} = {\Delta _{\rm{m}}}$ (i.e., the detuning frequencies of the mechanical and cavity modes are close to resonance) and the sideband resolved regime $\rm{Max} \left( {{\kappa _a},{\gamma _b}} \right) \ll {\Delta _{\rm{m}}}$ (i.e., the maximum dissipation rate is much smaller than the mechanical detuning) should hold simultaneously \cite{ref-2,ref-82,ref-83}. Under these validity conditions, we can safely apply the RWA to omit the anti-RWA terms $\delta a\delta b$ and $\delta {a^\dag }\delta {b^\dag }$, since they are strongly non-resonant; applying these terms to a state changes the total energy by an amount much larger than the coupling. Later, we verify numerically the rationality of the RWA applied here.

We now move to the RWA case. Keeping only the resonant terms $\delta a\delta {b^\dag }$ and $\delta {a^\dag }\delta b$ of the linearized QLEs.~(\ref{eq-29})-(\ref{eq-30}), we have
\begin{eqnarray}
\dot \nu  =  - {\rm M}\upsilon  + L{\upsilon _{{\rm{in}}}}, \label{eq-48}
\end{eqnarray}
where the quantum fluctuation vector is written as $\upsilon  = {\left( {\delta a,\delta b} \right)^T}$. The noise operator is written in a matrix form ${\upsilon _{{\rm{in}}}} = {\left( {\delta {a_{{\rm{in}}}},\delta {b_{{\rm{in}}}}} \right)^T}$. The damping operators in a matrix form read $L = {\rm{diag}}\left( {\sqrt {{\kappa _a}} ,\sqrt {{\gamma _b}} } \right)$. The coefficient matrix ${\rm M}$ in terms of system parameters takes the form
\begin{eqnarray}
{\rm M} = \left[ {\begin{array}{*{20}{c}}
{i{\Delta ''_{\rm{c}}} + \frac{{{\kappa _a}}}{2}}&{ - i{G_R}}\\
{ - i{G_R}}&{i{\Delta _{\rm{m}}} + \frac{{{\gamma _b}}}{2}}
\end{array}} \right], \label{eq-49}
\end{eqnarray}
The stability of the system requires that the real parts of all eigenvalues of the coefficient matrix ${\rm M}$ be positive, which can be analyzed using the Routh–Hurwitz criterion. Using the circuit QED parameters listed in Table~\ref{tab-1}, we set ${\Delta ''_{\rm{c}}} = {\Delta _{\rm{m}}} = \Delta $ and ${\kappa _a} = {\gamma _b} = \xi $. Solving the eigenvalue equation det$\left| {{\rm{M}} - \varepsilon {{\rm{I}}_2}} \right| = 0$, where ${{{\rm{I}}_2}}$ denotes the $2 \times 2$ identity matrix, we obtain two eigenvalues, ${\varepsilon _1} = i\left( {\Delta  - {G_R}} \right) + {\xi  \mathord{\left/{\vphantom {\xi  2}} \right.\kern-\nulldelimiterspace} 2}$ and ${\varepsilon _2} = i\left( {\Delta  + {G_R}} \right) + {\xi  \mathord{\left/{\vphantom {\xi  2}} \right. \kern-\nulldelimiterspace} 2}$. Their real parts satisfy ${\rm{Re}}\left[ {{\varepsilon  _1}} \right] = {\rm{Re}}\left[ {{\varepsilon_2}} \right] = {\xi  \mathord{\left/{\vphantom {\xi  2}} \right.\kern-\nulldelimiterspace} 2} > 0$, demonstrating that the steady-state stability condition is fulfilled for the parameters given in Table~\ref{tab-1}.

Similarly to the case without RWA, we find the solution to the linearized QLEs.~(\ref{eq-29})-(\ref{eq-30}) in the frequency domain as
\begin{eqnarray}
\upsilon \left( \omega  \right) = {\left[ {{\rm{M}} - i\omega {{\rm{I}}_2}} \right]^{ - 1}}L{\upsilon _{{\rm{in}}}}\left( \omega  \right). \label{eq-50}
\end{eqnarray}
Substituting Eq.~(\ref{eq-50}) into the input-output relation ${\upsilon _{{\rm{out}}}}\left( \omega  \right) + {\upsilon _{{\rm{in}}}}\left( \omega  \right) = {L}\upsilon \left( \omega  \right)$, we obtain
\begin{eqnarray}
{\upsilon _{{\rm{out}}}}\left( \omega  \right) = {\rm{X}}\left( \omega  \right){\upsilon _{{\rm{in}}}}\left( \omega  \right), \label{eq-51}
\end{eqnarray}
where the output field matrix ${\upsilon _{{\rm{out}}}}\left( \omega  \right)$ is the Fourier transform of ${\upsilon _{{\rm{out}}}} = {(\delta {a_{{\rm{out}}}},\delta {b_{{\rm{out}}}})^T}$. The scattering matrix for the symmetric optomechanical system can be written as
\begin{eqnarray}
{\rm{X}}\left( \omega  \right) = {L^T}{\left[ {{\rm{M}} - i\omega {{\rm{I}}_2}} \right]^{ - 1}}L - {{\rm{I}}_2}. \label{eq-52}
\end{eqnarray}

The spectrum of the output field is defined as
\begin{eqnarray}
{s_{{\rm{out}}}}\left( \omega  \right) = \int {d\omega '} \langle \upsilon _{{\rm{out}}}^\dag \left( {\omega '} \right){\upsilon _{{\rm{out}}}}\left( \omega  \right)\rangle.  \label{eq-53}
\end{eqnarray}
Substituting Eq.~(\ref{eq-51}) into Eq.~(\ref{eq-53}), we find
\begin{eqnarray}
{s_{{\rm{out}}}}\left( \omega  \right) = T\left( \omega  \right){s_{{\rm{in}}}}\left( \omega  \right),\label{eq-54}
\end{eqnarray}
where $T\left( \omega  \right)$ is the contribution arising from the input laser field. The specific forms of the scattering probability can be expressed as
\begin{eqnarray}
a \!\to \!b:T_a^b\left( \omega  \right) = {\left| {{\rm X}_a^b\left( \omega  \right)} \right|^2}; a \!\leftarrow \! b:T_b^a\left( \omega  \right) = {\left| {{\rm X}_b^a\left( \omega  \right)} \right|^2} \!\!.\label{eq-55}
\end{eqnarray}

In Sec.~{\rm \ref{section4}}-B, we use the parameter values for the QED circuit in the Table~\ref{tab-1} to carry out a numerical calculation to explore the conditions for the optimal reciprocal transmission of the incoming laser field.  The QED circuit in Table~\ref{tab-1} produces ${\Delta ''_{\rm{c}}} = {\Delta _{\rm{m}}} \approx 158{\kappa _a} = 158{\gamma _b}$. Here, we numerically verify the applicability of the RWA in this case. We here compare the scattering probability of a symmetric optomechanical system ${\rm{P}}_a^b = {\rm{P}}_b^a$ without RWA in Eqs.~(\ref{eq-46})-(\ref{eq-47}) and the one $T_a^b = T_b^a$ after RWA in Eq.~(\ref{eq-55}).

\begin{figure}[t]
\centering
\includegraphics[angle=0,width=0.468\textwidth]{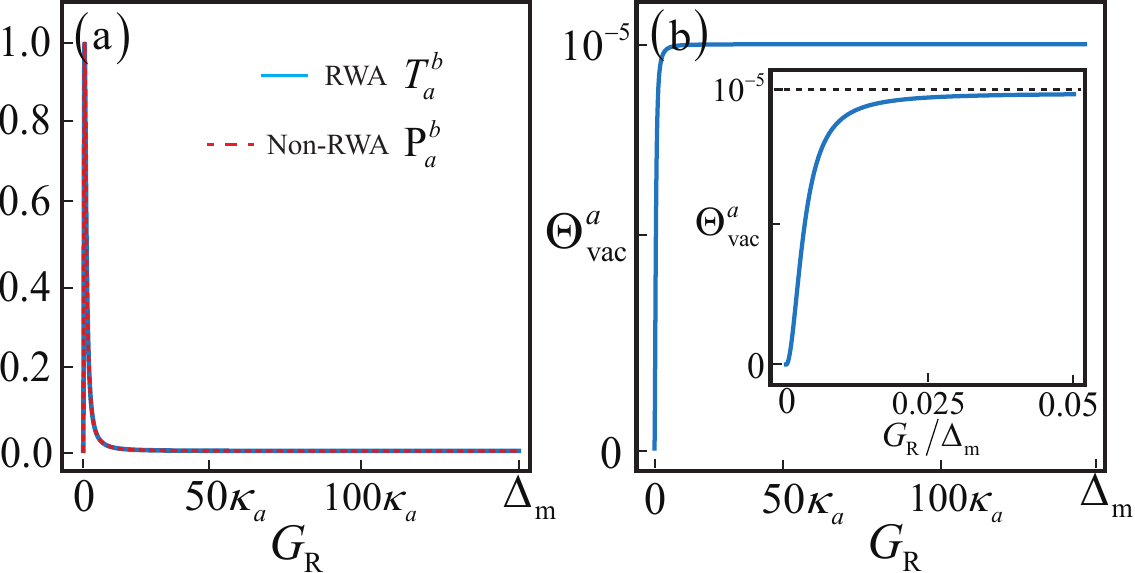}
\caption{$\left( \rm{a} \right)$ Scattering probabilities $T_a^b$ after RWA (solid blue line) and ${\rm P}_a^b$ without RWA (orange dashed line) as functions of the effective optomechanical coupling ${G_R} \in \left[ {0,{\Delta _{\rm{m}}}} \right]$. $\left( \rm{b} \right)$ Output spectral contribution from the incoming optical vacuum field $\Theta _{{\rm{vac}}}^a$ as a function of the effective optomechanical coupling ${G_R} \in \left[ {0,{\Delta _{\rm{m}}}} \right]$. The other parameters for $\left( \rm{a} \right)$ and $\left( \rm{b} \right)$ are chosen as $\omega  = {\Delta ''_{\rm{c}}} = {\Delta _{\rm{m}}} = 158{\kappa _a} = 158{\gamma _b}$.\label{Fig-5}}
\end{figure}

In Fig.~\ref{Fig-5}$\left( {\rm{a}} \right)$, we confirm that when the frequency of the input laser field satisfies the resonance condition $\omega  = {\Delta ''_{\rm{c}}} = {\Delta _{\rm{m}}}$ and sideband resolved regime $\rm{Max} \left( {{\kappa _a},{\gamma _b}} \right) \ll {\Delta _m}$ for any effective optomechanical coupling we have ${\rm{P}}_a^b \approx T_a^b$, that is, we can ignore the output spectrum resulting from the incoming optical vacuum field $\Theta _{{\rm{vac}}}^a$. Therefore, in Sec.~{\rm \ref{section4}}-B, we can safely omit the non-resonant terms $\delta a\delta b$ and $\delta {a^\dag }\delta {b^\dag }$ using the RWA. In addition, Fig.~\ref{Fig-5}$\left( \rm{b} \right)$ shows that with the enhancement of the effective optomechanical coupling ${{G_R}}$, the function $\Theta _{{\rm{vac}}}^a$ becomes more and more obvious, eventually approaching ${10^{ - 5}}$. After that, it exhibits robustness to ${{G_R}}$.

\begin{figure}[t]
\centering
\includegraphics[angle=0,width=0.36\textwidth]{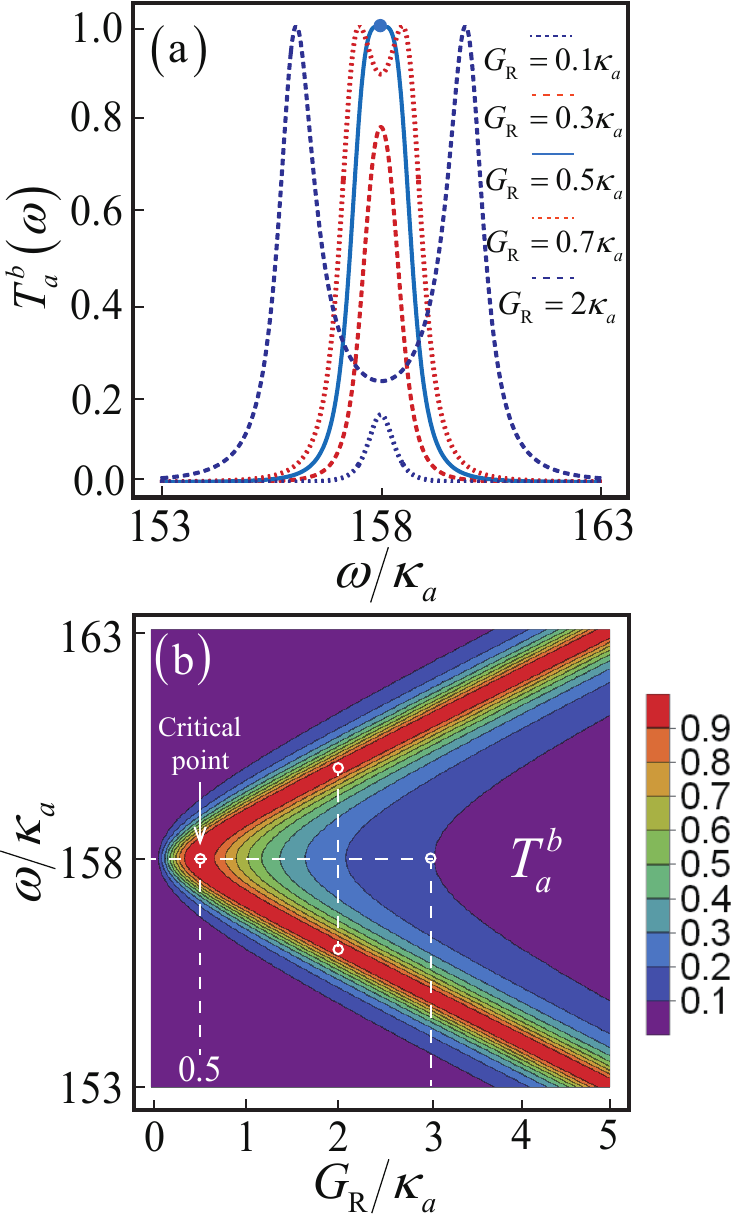}
\caption{The relation between the optimal reciprocal transmission of the incoming laser field and the effective optomechanical coupling ${{G_R}}$. $\left( \rm{a} \right)$ Scattering probability $T_a^b\left( \omega  \right)$ from the optical mode $a$ to the mechanical mode $b$ as a functions of frequency of the incoming singal $\omega $ (in units of ${{\kappa _a}}$) for different values of effective optomechanical coupling: ${G_R} = 0.1{\kappa _a}$ (purple dotted line), ${G_R} = 0.3{\kappa _a}$ (orange dashed line), ${G_R} = 0.5{\kappa _a}$ (solid blue line), ${G_R} = 0.7{\kappa _a}$ (orange dotted line), ${G_R} = 2{\kappa _a}$ (purple dashed line). $\left( \rm{b} \right)$ Scattering probability $T_a^b$ as a function of $\omega $ (in units of ${{\kappa _a}}$) and ${G_R}$ (in units of ${\kappa _a}$). The other parameters for $\left( \rm{a} \right)$ and $\left( \rm{b} \right)$ are set to ${\Delta ''_{\rm{c}}} = {\Delta _{\rm{m}}} = 158{\kappa _a} = 158{\gamma _b}$. \label{Fig-6}}
\end{figure}

\subsection{Critical points and optimal transmission condition}

In this part, to ensure that the maximum scattering probability of the input laser field is unity under the parameter values for the QED circuit in Table~\ref{tab-1}, we numerically calculate the scattering probability $T_a^b\left( \omega  \right)$ from the optical mode $a$ to the mechanical mode $b$ after RWA to show the optimal reciprocal transmission in the symmetric optomechanics, which can be widely applied to the optical antenna and lossless isotropic materials \cite{ref-84}. Here, the symmetric optomechanics demands that the effective optomechanical coupling ${{G_R}}$ is a real number. Although the symmetric optomechanical system provides the framework and serves as a baseline for the dynamical analysis in this work. It is worth mentioning that for $G$ to be a complex number, in particular, as demonstrated in our previous work \cite{ref-85}, in a three-mode optomechanical circulatory system the relative phase differences among distinct optomechanical couplings cannot be gauged away. These complex phase differences break time-reversal symmetry, leading to nonreciprocal transmission of an incident laser field when the phase difference satisfies $ \pm 0.5\pi  + 2n\pi $ for $n = Z$. This mechanism is not operative in a two-mode system—since the corresponding physical phase can be eliminated by an appropriate transformation of the reference frame—but it may become relevant in more general multimode optomechanical setting.

Now, we focus on the numerical evaluation of the scattering probability $T_a^b\left( \omega  \right)$ to show the optimal reciprocal transmission. The numerical results show the critical points and optimal condition for the observation of the optimal reciprocal transmission.

First, in Fig.~\ref{Fig-6}, for ${\kappa _a} = {\gamma _b} \ll {\Delta _{\rm{m}}}$, the scattering probability $T_a^b$ is investigated as a function of the incoming laser frequency $\omega $ and the effective optomechanical coupling ${{G_R}}$. Figure~\ref{Fig-6} shows that the maximum value of $T_a^b\left( \omega  \right)$ increases with ${G_R}$ in a weakly coupled region ${G_R} < 0.5{\kappa _a}$. In particular, the optimal reciprocal transmission $T_a^b\left( \omega  \right) = 1$ occurs for the first time at the strong-coupling critical point ${G_R} = 0.5{\kappa _a}$ when $\omega  = {\Delta ''_{\rm{c}}} = {\Delta _{\rm{m}}}$, that is, the incoming laser frequency resonates with both the modified detuning frequency of the optical cavity field ${\Delta ''_{\rm{c}}}$ and the mechanical detuning frequency ${\Delta _{\rm{m}}}$. In the strong coupling region ${G_R} > 0.5{\kappa _a}$, on the other hand, with enhancement of ${G_R}$, the single peak splits into two peaks, and the distance between the two symmetrical peaks become wider as ${{G_R}}$ further increases. As ${{G_R}}$ is further strengthened, two symmetrical peaks for the optimal reciprocal transmission always appear with the resonant frequency $\omega  = {\Delta ''_{\rm{c}}} = {\Delta _{\rm{m}}}$ being the axis of symmetry.

\begin{figure}[t]
\centering
\includegraphics[angle=0,width=0.36\textwidth]{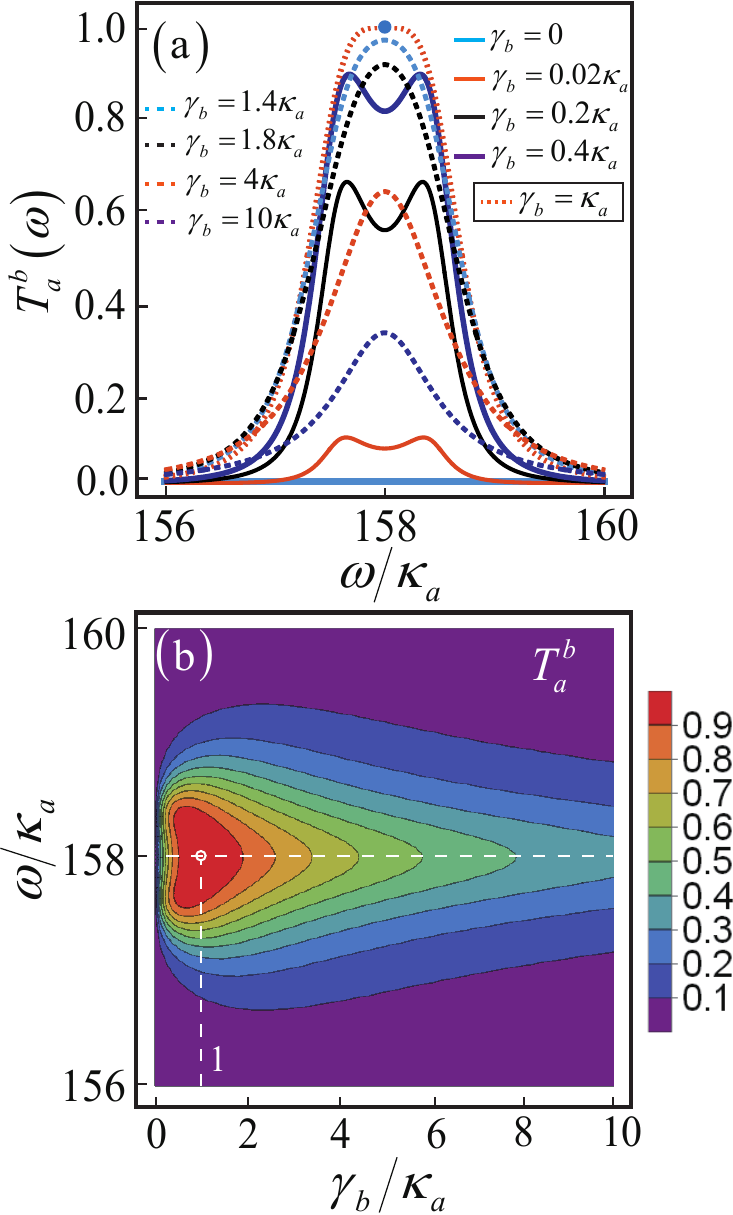}
\caption{The relation between the optimal reciprocal transmission of the incoming laser field and the mechanical damping rate ${{\gamma _b}}$. $\left( \rm{a} \right)$ The scattering probability $T_a^b\left( \omega  \right)$ as functions of the frequency of the incoming signal $\omega $ (in units of ${{\kappa _a}}$) for different mechanical damping rates: ${\gamma _b} = 0$ (blue solid line), ${\gamma _b} = 0.02{\kappa _a}$ (orange solid line), ${\gamma _b} = 0.2{\kappa _a}$ (black solid line), ${\gamma _b} = 0.4{\kappa _a}$ (purple solid line), ${\gamma _b} = {\kappa _a}$ (orange dotted line),  ${\gamma _b} = 1.4{\kappa _a}$ (blue dashed line), ${\gamma _b} = 1.8{\kappa _a}$ (black dashed line), ${\gamma _b} = 4{\kappa _a}$ (orange dashed line), ${\gamma _b} = 10{\kappa _a}$ (purple dashed line). The other parameters are set to ${\Delta ''_{\rm{c}}} = {\Delta _{\rm{m}}} = 158{\kappa _a}$ and ${{G_R} = 0.5{\kappa _a}}$. $\left( \rm{b} \right)$ Scattering probability $T_a^b$ as a function of $\omega $ (in units of ${{\kappa _a}}$) and ${\gamma _b}$ (in units of ${{\kappa _a}}$) with the other parameters are set to ${\Delta ''_{\rm{c}}} = {\Delta _{\rm{m}}} = 158{\kappa _a}$ and ${{G_R} = 0.5{\kappa _a}}$. \label{Fig-7}}
\end{figure}

\begin{figure}[h]
\centering
\includegraphics[angle=0,width=0.38\textwidth]{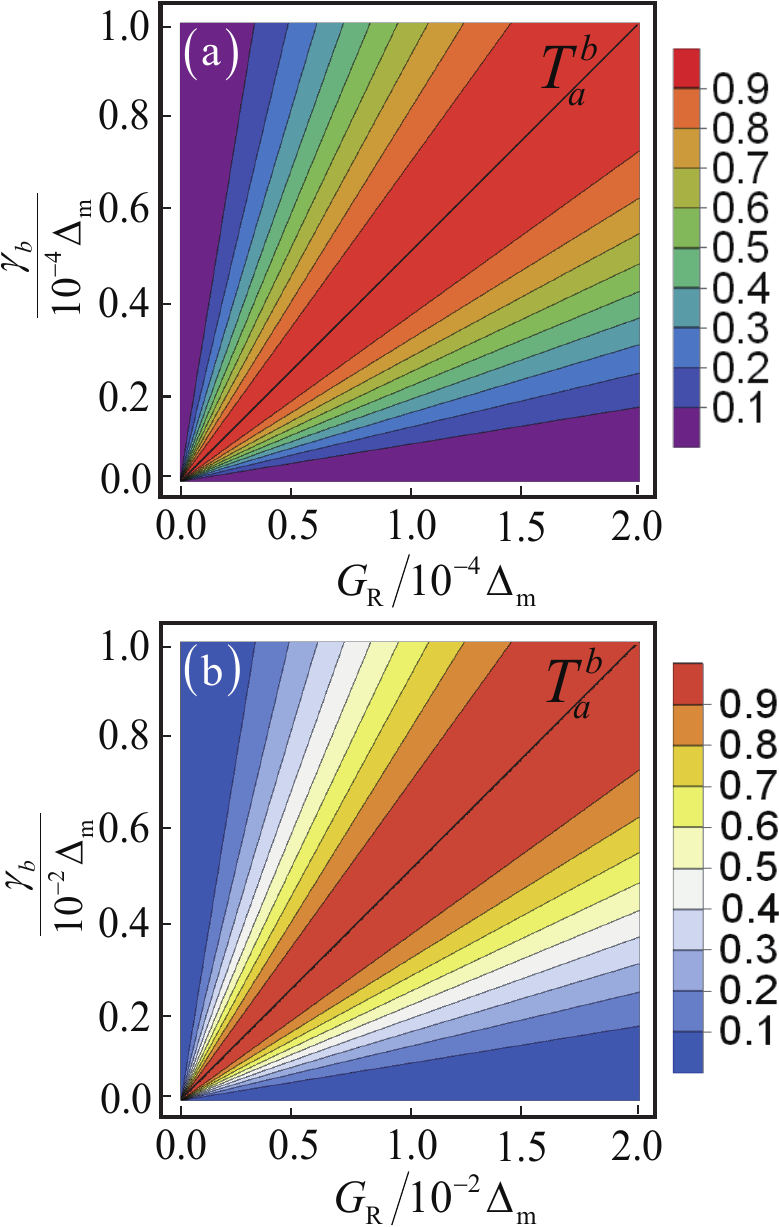}
\caption{Scattering probability $T_a^b$ as a function of the effective optomechanical coupling ${{G_R}}$ and the mechanical damping rate ${{\gamma _b}}$ with the other parameters are set to $\omega  = {\Delta ''_{\rm{c}}} = {\Delta _{\rm{m}}} \gg {\kappa _a} = {\gamma _b}$ and ${G_R} = 0.5{\kappa _a}$. $\left( \rm{a} \right)$ Case of weak dissipation: ${\gamma _b}$ approaches ${10^{ - 4}}{\Delta _{\rm{m}}}$. $\left( \rm{b} \right)$ Case of strong dissipation: ${\gamma _b}$ approaches ${10^{ - 2}}{\Delta _{\rm{m}}}$. \label{Fig-8}}
\end{figure}

\begin{figure}[t]
\centering
\includegraphics[angle=0,width=0.36\textwidth]{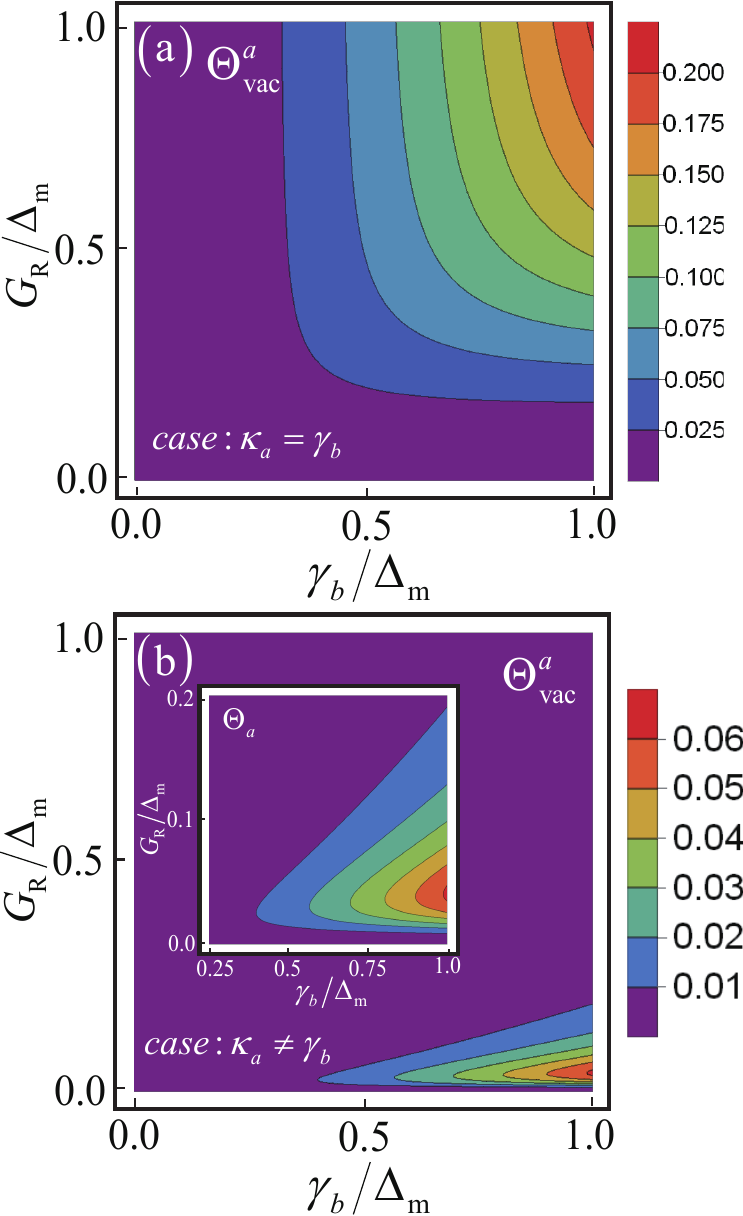}
\caption{Output spectral contribution from the incoming optical vacuum field $\Theta _{{\rm{vac}}}^a$ as a function of the damping ${\gamma _b}$ of the mechanical oscillator and the effective optomechanical coupling ${G_R}$. $\left( \rm{a} \right)$ Case of dissipative equilibrium: ${\kappa _a} = {\gamma _b} \in \left( {0,{\Delta _{\rm{m}}}} \right)$. The other parameters are chosen as $\omega  = {\Delta ''_{\rm{c}}} = {\Delta _{\rm{m}}}$ and ${G_R} \in \left( {0,{\Delta _{\rm{m}}}} \right)$. $\left( \rm{b} \right)$ Case of dissipative non-equilibrium: ${\kappa _a} \ne {\gamma _b} \in \left( {0,{\Delta _{\rm{m}}}} \right)$ with ${\kappa _a}=1$. The other parameters are set to $\omega  = {\Delta ''_{\rm{c}}} = {\Delta _{\rm{m}}} = 158{\kappa _a}$ and ${G_R} \in \left( {0,{\Delta _{\rm{m}}}} \right)$. \label{Fig-9}}
\end{figure}

\begin{figure*}[t]
\centering
\includegraphics[angle=0,width=0.8\textwidth]{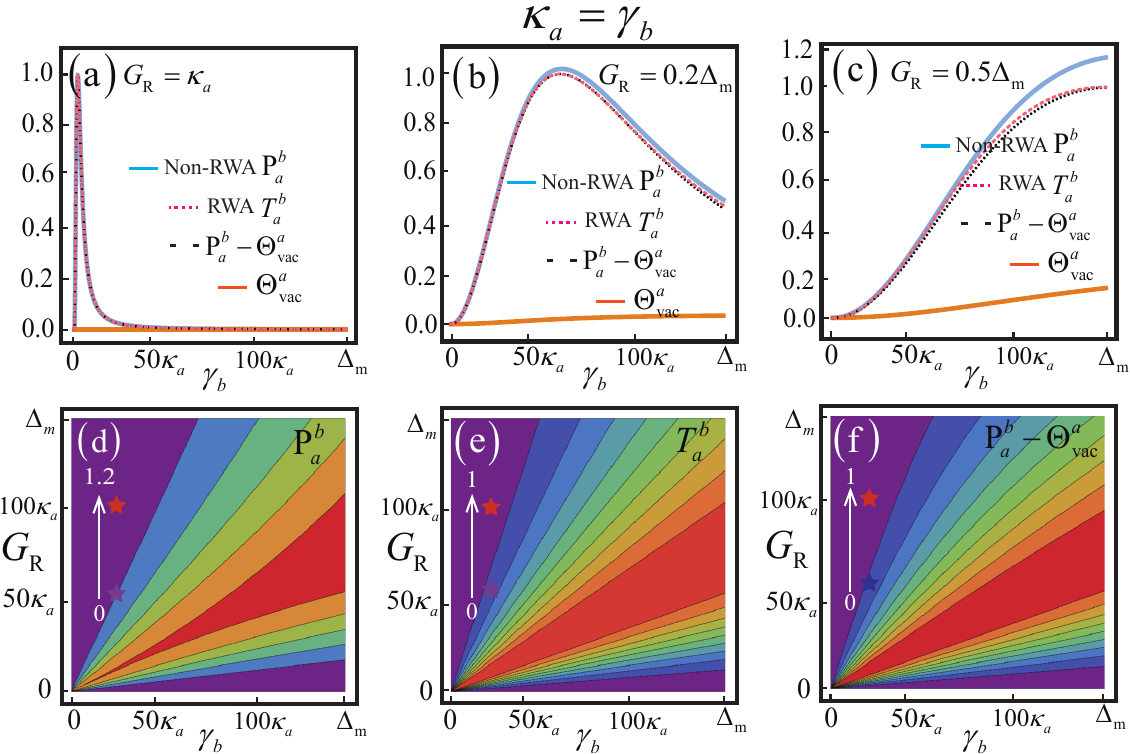}
\caption{Case of dissipative equilibrium: ${\kappa _a} = {\gamma _b} \in \left( {0,{\Delta _{\rm{m}}}} \right)$. $\left( {\rm{a}} \right)$-$\left( \rm{c} \right)$: We compare the scattering probabilities ${\rm{P}}_a^b$ before RWA and $T_a^b$ after it as functions of the mechanical damping ${{\gamma _b}}$ for different effective optomechanical coupling ${G_R}$: $\left( \rm{a} \right)$ ${G_R} = {\kappa _a}$, $\left( \rm{b} \right)$ ${G_R} = 0.2{\Delta _{\rm{m}}}$, $\left( \rm{c} \right)$ ${G_R} = 0.5{\Delta _{\rm{m}}}$. $\left( {\rm{d}} \right)$-$\left( {\rm{f}} \right)$: Scattering probabilities ${\rm{P}}_a^b$, $T_a^b$, and ${\rm{P}}_a^b - \Theta _{{\rm{vac}}}^a$ as functions of ${{\gamma _b}}$ and different values of ${G_R}$. The other parameters are set to $\omega  = {\Delta ''_{\rm{c}}} = {\Delta _{\rm{m}}}$ in $\left( {\rm{a}} \right)$-$\left( f \right)$. \label{Fig-10}}
\end{figure*}

\begin{figure*}[t]
\centering
\includegraphics[angle=0,width=0.8\textwidth]{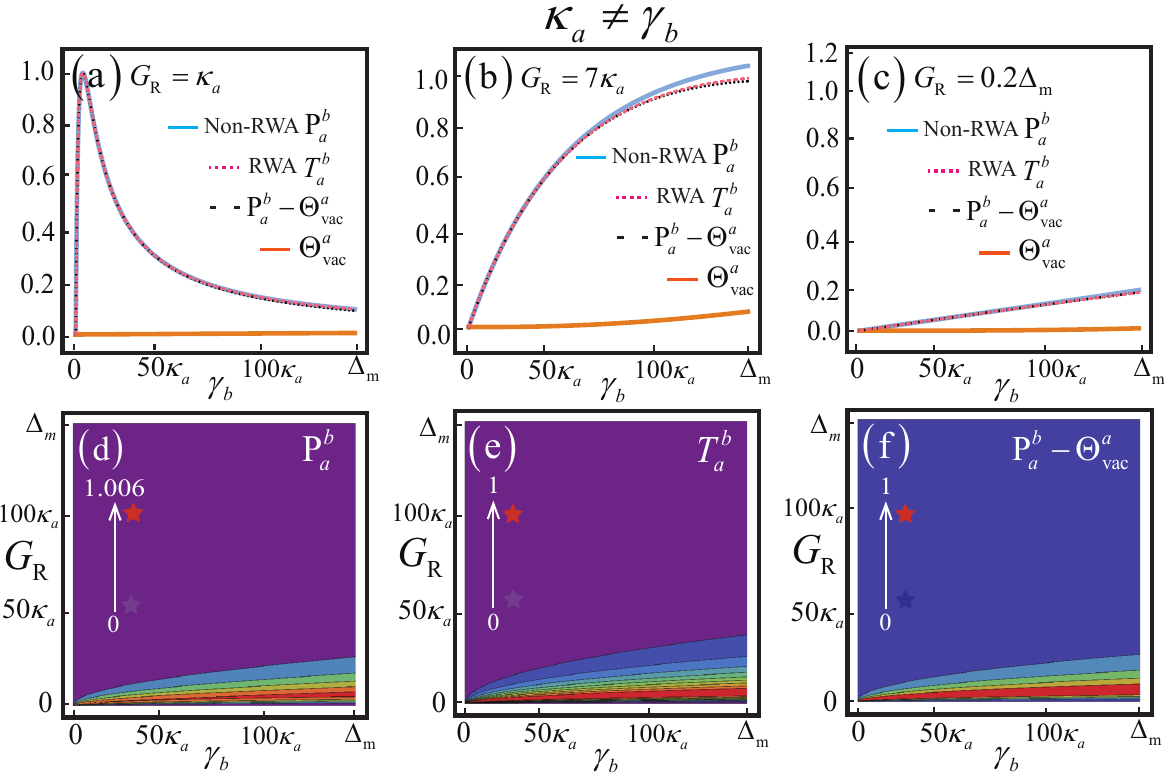}
\caption{Case of dissipative non-equilibrium: ${\kappa _a} \ne {\gamma _b} \in \left( {0,{\Delta _{\rm{m}}}} \right)$ with ${\kappa _a}=1$. $\left( {\rm{a}} \right)$-$\left( c \right)$: We compare the scattering probabilities ${\rm P}_a^b\left( {{\gamma _b}} \right)$ before RWA and $T_a^b\left( {{\gamma _b}} \right)$ after it as functions of ${{\gamma _b}}$ for different values of ${G_R}$: $\left( a \right)$ ${G_R} = {\kappa _a}$, $\left( b \right)$ ${G_R} = 7{\kappa_a}$, $\left( c \right)$ ${G_R} = 0.2{\Delta _{\rm{m}}}$. $\left( {\rm{d}} \right)$-$\left( {\rm{f}} \right)$: Scattering probabilities ${\rm{P}}_a^b$, $T_a^b$, and ${\rm{P}}_a^b - \Theta _{{\rm{vac}}}^a$ as functions of ${{\gamma _b}}$ and different values of ${G_R}$. The other parameters are set to $\omega  = {\Delta ''_{\rm{c}}} = {\Delta _{\rm{m}}} = 158{\kappa _a}$ in $\left( {\rm{a}} \right)$-$\left( f \right)$. \label{Fig-11}}
\end{figure*}

\begin{table*}[htbp]
	\centering
    \setlength{\abovecaptionskip}{0pt}
    \setlength{\belowcaptionskip}{8pt}
    \renewcommand{\arraystretch}{1}
    \setlength\tabcolsep{3.6mm}
	\caption{Summary of the results of our study on optimal reciprocal transmission: $\left( a \right)$ optomechanical classification according to coupling type. $\left( b \right)$ The relation between the optimal reciprocal transmission of the incoming laser field and the effective optomechanical coupling ${{G_R}}$. The other parameters are set to $\omega  = {\Delta ''_{\rm{c}}} = {\Delta _{\rm{m}}} \gg {\kappa _a} = {\gamma _b}$. $\left( c \right)$ The relation between the optimal reciprocal transmission of the incoming laser field and the mechanical damping rate ${{\gamma _b}}$. The other parameters are set to ${\Delta ''_{\rm{c}}} = {\Delta _{\rm{m}}} \gg {\kappa _a}$ and ${G_R} = 0.5{\kappa _a}$.}
	\label{tab-3}
    \begin{tabular}{ccccc ccccc}
		\hline\hline\noalign{\smallskip}
         $\left( a \right)$ Optomechanical type & Coupling & Dissipation & Frequency & Transmission Behavior\\
         \hline\noalign{\smallskip}
         Multimode circulatory optomechanics & $G \in C$; \ $\left| G \right| = 0.5{\kappa _a}$ & ${\gamma _b} = {\kappa _a}$ & ${\Delta ''_{\rm{c}}} = {\Delta _{\rm{m}}}$& Non-reciprocity \cite{ref-85,ref-86,ref-87}\\
         Symmetrical optomechanics & ${G_R} \in R$; \ ${G_R} = 0.5{\kappa _a}$ & ${\gamma _b} = {\kappa _a}$ & ${\Delta ''_{\rm{c}}} = {\Delta _{\rm{m}}}$& Reciprocity\\
         \hline\hline\noalign{\smallskip}
         $\left( b \right)$ Real coupling & $\rm{Max} \ T_a^b\left( {{G_R}} \right)$ & Peak Number & Energy gap & $T_a^b\left( {{G_R}} \right) = 1$ points\\
         \hline\noalign{\smallskip}
          ${G_R} < 0.5{\kappa _a}$ & ${G_R} \uparrow T_a^b\left( {{G_R}} \right) \uparrow $ & Single & N/A & Zero\\
          ${G_R} = 0.5{\kappa _a}$ & $T_a^b\left( {{G_R}} \right) \equiv 1$ & Single & N/A & One\\
          ${G_R} > 0.5{\kappa _a}$ & $T_a^b\left( {{G_R}} \right) = 1$ & Double & ${G_R} \uparrow \rm{Gap} \uparrow $ & Two\\
         \hline\hline\noalign{\smallskip}
           $\left( c \right)$ Dissipation & $\rm{Max} \ T_a^b\left( {{\gamma _b}} \right)$ & Peak Number & Energy gap & $T_a^b\left( {{\gamma _b}} \right) = 1$ points\\
           \hline\noalign{\smallskip}
           ${\gamma _b} < {\kappa _a}$ & ${\gamma _b} \uparrow T_a^b\left( {{\gamma _b}} \right) \uparrow $ & Double & ${\gamma _b} \uparrow \rm{Gap} \uparrow $ & Zero\\
            ${\gamma _b} = {\kappa _a}$ & $T_a^b\left( {{\gamma _b}} \right) \equiv 1$ & Single & N/A & One\\
             ${\gamma _b} > {\kappa _a}$ & ${\gamma _b} \uparrow T_a^b\left( {{\gamma _b}} \right) \downarrow $ & Single & N/A & Zero\\
             \hline\hline\noalign{\smallskip}
         \end{tabular}
\end{table*}

Then, we explore the influence of the environment on the optimal reciprocal transmission. In Fig.~\ref{Fig-7}, we plot the scattering probability $T_a^b$ as a function of the incoming laser frequency $\omega $ and the mechanical damping rate ${{\gamma _b}}$ with the other parameters are set to ${\Delta ''_{\rm{c}}} = {\Delta _{\rm{m}}} = 158{\kappa _a}$ and ${{G_R} = 0.5{\kappa _a}}$. It is shown that for ${\gamma _b} < {\kappa _a}$, the maximum value of $T_a^b\left( \omega  \right)$ increases as ${\gamma _b}$ increases. In particular, we achieve the optimal reciprocal transmission when the dissipative balanced condition ${\gamma _b} = {\kappa _a}$ is satisfied. For ${\gamma _b} > {\kappa _a}$, as ${{\gamma _b}}$ further increases, the maximum value of $T_a^b\left( \omega  \right)$ gradually decreases, too. In addition, we observe that for ${\gamma _b} < {\kappa _a}$, the scattering probability $T_a^b\left( \omega  \right)$ always presents a double peak with the resonant frequency $\omega  = {\Delta ''_{\rm{c}}} = {\Delta _{\rm{m}}}$ being the axis of symmetry. With the further increase of ${\gamma _b}$, the scattering probability $T_a^b\left( \omega  \right)$ presents a single peak at the resonant frequency $\omega  = {\Delta ''_{\rm{c}}} = {\Delta _{\rm{m}}}$. Here, in contrast to Fig.~\ref{Fig-6}, where $T_a^b$ varies with ${{G_R}}$, we can see that $T_a^b$ varies with ${{\gamma _b}}$ in the opposite process. In short, the effective optomechanical coupling ${G_R}$ promotes the splitting of the energy levels whereas the mechanical damping rate ${\gamma _b}$ suppresses it.

Considering all these factors, in Fig.~\ref{Fig-8}, we plot the scattering probability $T_a^b$ as a function of ${{G_R}}$ and ${{\gamma _b}}$ with the other parameters set to $\omega  = {\Delta ''_{\rm{c}}} = {\Delta _{\rm{m}}} \gg {\kappa _a} = {\gamma _b}$. We are surprised to find that as long as the condition ${G_R} = 0.5{\kappa _a}$ is satisfied, the optimal reciprocal transmission always exists from an undamped optomechanical system to a strongly damped optomechanical system. Now, we analytically verify our numerical observation. In our case, if the parameters satisfy $\omega  = {\Delta ''_{\rm{c}}} = {\Delta _{\rm{m}}} = \Delta  \gg {\kappa _a} = {\gamma _b} = \xi $, the scattering matrix~(\ref{eq-52}) can be cast into the following form:
\begin{eqnarray}
{\rm X}\left( {{G_R},\xi } \right) = \left[ {\begin{array}{*{20}{c}}
{\frac{{0.5{\xi ^2}}}{{0.25{\xi ^2} + G_R^2}} - 1}&{I\frac{{\xi {G_R}}}{{0.25{\xi ^2} + G_R^2}}}\\
{I\frac{{\xi {G_R}}}{{0.25{\xi ^2} + G_R^2}}}&{\frac{{0.5{\xi ^2}}}{{0.25{\xi ^2} + G_R^2}} - 1}
\end{array}} \right].\label{eq-56}
\end{eqnarray}
The scattering probability $T\left( \omega  \right)$ is given by $T\left( \omega  \right) = {\left| {{\rm X}\left( {{G_R},\xi } \right)} \right|^2}$. For ${G_R} = 0.5\xi $, the scattering matrix~(\ref{eq-56}) reduces to ${\rm X}\left( {{G_R} = 0.5\xi } \right) = \left[ {\left\{ {0,I} \right\},\{ I,0\} } \right]$, and thus we find $T_a^b\left( {{G_R} = 0.5\xi } \right) = T_b^a\left( {{G_R} = 0.5\xi } \right) = 1$.

Finally, we summarize the results of our study on the optimal reciprocal transmission as follows; see Table~\ref{tab-3}. The optimal reciprocal transmission needs to satisfy both of the following conditions: the incoming laser frequency $\omega $ resonates with both the modified detuning frequency of the optical cavity field ${\Delta ''_{\rm{c}}}$ and the mechanical detuning frequency ${\Delta _{\rm{m}}}$; the decay of the optical cavity ${\kappa _a}$ is equal to the mechanical damping rate ${\gamma _b}$; the optomechanical coupling ${{G_R}}$ is equal to ${0.5{\kappa _a}}$. In addition, it is worth mentioning that for ${G_R} > 0.5{\kappa _a}$ and ${\gamma _b} = {\kappa _a}$, two symmetrical peaks for optimal reciprocal transmission always appear with the resonant frequency $\omega  = {\Delta ''_{\rm{c}}} = {\Delta _{\rm{m}}}$ being the axis of symmetry. When ${G_R} > 3{\kappa _a}$ and ${\gamma _b} = {\kappa _a}$, the photon blocking effect is exhibited at the frequency resonance $\omega  = {\Delta ''_{\rm{c}}} = {\Delta _{\rm{m}}}$, as we observe $T_a^b\left( {{\Delta _{\rm{m}}}} \right) \approx 0$.

\subsection{Non-RWA for dissipative equilibrium and non-equilibrium symmetric optomechanics}

In this part, we consider the case of strong dissipation and ultrastrong coupling, in which the sideband resolved regime ${\rm{max}}\left( {{\kappa _{\rm{a}}},{\gamma _{\rm{b}}}} \right) \ll {\Delta _{\rm{m}}}$ is no longer met. Setting $\omega  = {\Delta ''_{\rm{c}}} = {\Delta _{\rm{m}}}$ and ${\rm{max}}\left( {{\kappa _{\rm{a}}},{\gamma _{\rm{b}}}} \right) \in \left( {0,{\Delta _{\rm{m}}}} \right)$, we point out the contribution of the incoming optical vacuum field to the output spectrum $\Theta _{{\rm{vac}}}^a$, which is an effect of the anti-RWA terms, cannot always be neglected.

In Fig.~\ref{Fig-9}, we plot $\Theta _{{\rm{vac}}}^a$ as a function of the damping of the mechanical oscillator ${\gamma _b}$ and the effective optomechanical coupling ${G_R}$. Figure~\ref{Fig-9}$\left( \rm{a} \right)$ shows a case of dissipative equilibrium. We find that in the case of strong dissipation ${\gamma _b}$ approaching ${\Delta _{\rm{m}}}$, the quantity $\Theta _{{\rm{vac}}}^a$ continues to increase as ${G_R}$ grows. In contrast to this, we give a case of dissipative non-equilibrium in Fig.~\ref{Fig-9}$\left( \rm{b} \right)$, in which there is a crossover region. When ${\gamma _b}$ approaches ${\Delta _{\rm{m}}}$, with the increase of ${{G_R}}$, the quantity $\Theta _{{\rm{vac}}}^a$ first increases, and then decreases.

In the dissipative equilibrium case ${\kappa _a} = {\gamma _b} \in \left( {0,{\Delta _{\rm{m}}}} \right)$, in Fig.~\ref{Fig-10}$\left( {\rm{a}} \right)$-$\left( {\rm{c}} \right)$, we compare the scattering probabilities ${\rm{P}}_a^b$ before RWA and $T_a^b$ after it as functions of ${{\gamma _b}}$ for different values of ${G_R}$. To be specific, when ${G_R} = {\kappa _a}$, we can see that ${\rm{P}}_a^b \approx T_a^b$. It is also shown that for ${G_R} = {\kappa _a}$ as the increase of ${\gamma _b}$, the scattering probabilities initially increases and then decreases. When ${G_R} = 0.2{\Delta _{\rm{m}}}$ and ${G_R} = 0.5{\Delta _{\rm{m}}}$, we have ${\rm{P}}_a^b > T_a^b$. We numerically find that the approximate relation ${\rm{P}}_a^b - \Theta _{{\rm{vac}}}^a \approx T_a^b$ always hold, which means that the difference between the non-RWA and the RWA cases is due to the contribution of the input vacuum field, $\Theta _{{\rm{vac}}}^a$. We also find that for ${G_R} = 0.5{\Delta _{\rm{m}}}$ with the increase of ${\gamma _b}$, the scattering probabilities continues to increase. In Fig.~\ref{Fig-10}$\left( \rm{d} \right)$-$\left( \rm{f} \right)$, we plot the scattering probabilities ${\rm{P}}_a^b$, $T_a^b$, and ${\rm{P}}_a^b - \Theta _{{\rm{vac}}}^a$ as functions of ${{\gamma _b}}$ and different values of ${G_R}$ with parameters $\omega  = {\Delta ''_{\rm{c}}} = {\Delta _{\rm{m}}}$ and ${\kappa _a} = {\gamma _b} \in \left( {0,{\Delta _{\rm{m}}}} \right)$. We find that the result in Fig.~\ref{Fig-10}$\left( {\rm{a}} \right)$-$\left( {\rm{f}} \right)$ is consistent with that in Fig.~\ref{Fig-9}$\left( \rm{a} \right)$.

Similarly, in the dissipative non-equilibrium case: ${\kappa _a} \ne {\gamma _b} \in \left( {0,{\Delta _{\rm{m}}}} \right)$ with ${\kappa _a}=1$, we also numerically find that the approximate relation ${\rm{P}}_a^b - \Theta _{{\rm{vac}}}^a \approx T_a^b$ always hold. We show that the result in Fig.~\ref{Fig-11} is consistent with that in Fig.~\ref{Fig-9}$\left( b \right)$. The significant difference between Fig.~\ref{Fig-10}$\left( {\rm{b}} \right)$ and Fig.~\ref{Fig-11}$\left( {\rm{c}} \right)$ in the ultrastrong region is that in the case of dissipative equilibrium, the scattering probabilities first increases rapidly and then decreases with the increase of ${\gamma _b}$, while in the case of dissipative non-equilibrium, the scattering probabilities continues to grow very slowly with the increase of ${\gamma _b}$.

In summary, we have discussed the effect of dissipation on laser signal transmission in a symmetrical optomechanical system. More importantly, we modify the expression of scattering probability under the non-RWA. Further, we numerically find that the approximate relation ${\rm{P}}_a^b - \Theta _{{\rm{vac}}}^a \approx T_a^b$ always hold.

\section{Summary and prospect}\label{section6}

In summary, based on an enhanced cross-Kerr circuit-QED platform, we have proposed a reliable scheme to achieve controllable enhancement of optomechanical coupling in the few-photon regime via dual coherent laser driving. Moreover, we introduced an effective symmetric optomechanical system, in which the quantum fluctuation dynamics of photons and phonons take analogous forms. Within this symmetric framework, we demonstrated optimal reciprocal transport and identified the boundary of the optomechanical strong-coupling regime by analyzing the critical behavior of the optimal laser-field transmission. Furthermore, we compared the scattering properties of the laser field in dissipative equilibrium and nonequilibrium symmetric optomechanical systems, both before and after applying the rotating-wave approximation. Our results provide a theoretical foundation for exploring a broad class of strongly coupled optomechanical quantum devices. 

This work also reveals how the form of optomechanical coupling and the type of dissipation affect the scattering characteristics of the input laser signal in an open quantum system. Further, we can explore the influence of optomechanical coupling and dissipation on optomechanical-induced transparency \cite{ref-92,ref-93,ref-94}, negative entanglement \cite{ref-95,ref-96,ref-97,ref-Shang044048}, as well as non-Markovian \cite{ref-Sun2504.09695,ref-Shen2503.21739} and non-Hermitian \cite{ref-Zhang063702, ref-Ning250400617} dynamics in the future.

\section*{Acknowledgements}
\noindent C. Shang is grateful to Naomichi Hatano for fruitful discussions and for carefully proofreading the manuscript. C. Shang also thanks Yue-Zhou Li and Hongchao Li for their valuable comments. H. Z. Shen was supported by the Science and Technology Development Plan Project of Jilin Province (Grant No.~20250102007JC), and the National Natural Science Foundation of China under Grant No.~12274064. C. Shang acknowledges financial support from the China Scholarship Council, the Japanese Government (Monbukagakusho-MEXT) Scholarship (Grant No.~211501), the RIKEN Junior Research Associate Program, and the Hakubi Projects of RIKEN. Additionally, we sincerely thank the referees for their careful review and constructive comments, which have significantly improved our work.

\section*{Conflict of Interest}
\noindent The authors declare no conflict of interest.

\section*{Data Availability Statement}
\noindent The data that support the findings of this study are available from the corresponding authors C. Shang and H. Z. Shen upon reasonable request.

\section*{Keywords}
\noindent few-photon regime, cross-Kerr nonlinearity, controllable optomechanical coupling, symmetric optomechanics, dissipative dynamics

\appendix
\section{Details of the derivation of Eq.~(\ref{eq-1})} \label{appendix A}
Here, we show the origin of an original optomechanical setup $H_{\rm{O}}$ from a decoupled cavity optomechanics \cite{ref-41,ref-42,ref-43,ref-44}. To begin with, let us consider an optomechanical system without interactions. Considering the limitation of optomechanical experiments, we set the Hamiltonian for an uncoupled optomechanical system in the simple form
\begin{eqnarray}
{H_{\rm A}} = \hbar {\omega _{\rm{c}}}{a^\dag }a + \hbar {\omega _{\rm{m}}}{b^\dag }b, \label{eq-100}
\end{eqnarray}
where ${a^\dag }$ $\left( a \right)$ and ${b^\dag }$ $\left( b \right)$ are the creation (annihilation) operators of the optical and mechanical modes with the corresponding resonance frequencies ${\omega _{\rm{c}}}$ and ${\omega _{\rm{m}}}$, respectively.

Then we assume that the movable mirror works with a slight mechanical motion. Thus, the resonance frequency of the optical cavity is modulated by the mechanical amplitude. In the present paper, we keep only the contributions of the linear and first nonlinear terms in the Taylor expansion:
\begin{eqnarray}
{\omega _{\rm{c}}}\left( x \right) = {\omega _{\rm{c}}} + x\frac{{\partial {\omega _{\rm{c}}}\left( x \right)}}{{\partial x}} + \frac{1}{2}{x^2}\frac{{{\partial ^2}{\omega _{\rm{c}}}\left( x \right)}}{{\partial {x^2}}} +  \cdots  \cdots. \label{eq-101}
\end{eqnarray}
Here, $x = {x_{{\rm{ZPF}}}}\left( {{b^\dag } + b} \right)$, where ${x_{{\rm{ZPF}}}}$ is the amplitude of the zero-point fluctuation of the mechanical oscillator, while ${{ - \partial {\omega _c}\left( x \right)} \mathord{\left/{\vphantom {{ - \partial {\omega _c}\left( x \right)} {\partial x}}} \right.\kern-\nulldelimiterspace} {\partial x}}$ is the shift of the optical frequency per displacement. Now, we define ${g_0}$ and $\chi' $ as follows:
\begin{eqnarray}
{g_{\rm{0}}} =  - {x_{{\rm{ZPF}}}}\frac{{\partial {\omega _{\rm{c}}}\left( x \right)}}{{\partial x}}, \quad {\chi'}  = x_{{\rm{ZPF}}}^2\frac{{{\partial ^2}{\omega _{\rm{c}}}\left( x \right)}}{{\partial {x^2}}}, \label{eq-102}
\end{eqnarray}
where ${g_0}$ and $\chi' $ are often known as the single-photon optomechanical coupling and the original cross-Kerr coupling, respectively. Hence, Eq.~(\ref{eq-101}) can be rewritten as
\begin{eqnarray}
{\omega _{\rm{c}}}\left( x \right) = {\omega _{\rm{c}}} - {g_{\rm{0}}}\left( {{b^\dag } + b} \right) + \frac{1}{2}\chi {\left( {{b^\dag } + b} \right)^2} + {\rm O}\left( {{x^3}} \right). \label{eq-103}
\end{eqnarray}
We further simplify Eq.~(\ref{eq-103}) by invoking the rotating-wave approximation (RWA), in which we omit the non-resonant terms ${b^\dag }{b^\dag }$ and $bb$ in the sideband resolved and red detuned regimes \cite{ref-79,ref-80}. Substituting the above result into Eq.~(\ref{eq-100}), we obtain Eq.~(\ref{eq-1}) in the main text, namely, {$H_{{\rm{O}}} = \hbar {\omega _{\rm{c}}}{a^\dag }a + \hbar {\omega _{\rm{m}}}{b^\dag }b - \hbar {g_{\rm{0}}}{a^\dag }a ( {{b^\dag } + b} ){\rm{ + }}\hbar \chi' {a^\dag }a{b^\dag }b$}.

Physically, the radiation-pressure interaction ${g_0}$ represents the frequency shift of the optical cavity field, which depends on the displacement of the mechanical resonator from the equilibrium position. The original cross-Kerr interaction $\chi' $ is interpreted as a change in the reflective index of the optical cavity field, which depends on the number of mechanical phonons.

\section{Details of the derivation of Eq.~(\ref{eq-9}) and ${\beta _{\rm{s}}}$} \label{appendix B}
In this Appendix, we derive the nonlinear quantum Langevin equations that the final enhanced Hamiltonian ${H_{\rm{T}}}$ satisfies \cite{ref-57}. Let us focus on the mechanical mode as an example; Eq.~(\ref{eq-8}) for the optical cavity mode is derived similarly. The Heisenberg equations of motion for the system operator $b$ and its corresponding reservoir operators ${\Gamma _{b{\rm{k}}}}$ are given by
\begin{eqnarray}
 \!\!\!\!\!  \dot b \!\!\!&=&\!\!\!  - i{\Delta _{\rm{m}}}b \!+\! i{g_{\rm{s}}}{a^\dag }a \!-\! i\left( {{\Omega _b} \!-\! {\beta _{\rm{s}}}{\Delta _{\rm{m}}}} \right) \!-\! i\sum\limits_{\rm{k}} {{g_{{\rm{bk}}}}{\Gamma _{{\rm{bk}}}}},  \label{eq-104}\\
\!\!\!\!\!  {{\dot \Gamma }_{b{\rm{k}}}} \!\!\!&=&\!\!\!  - i{\omega _{b{\rm{k}}}}{\Gamma _{b{\rm{k}}}} - i{g_{b{\rm{k}}}}\left( {b - \beta_{\rm{s}} } \right).\label{eq-105}
\end{eqnarray}
We are interested in a closed equation for the system operator $b$. Equation~(\ref{eq-105}) for the reservoir operator ${\Gamma _{b{\rm{k}}}}$ can be formally integrated to yield
\begin{eqnarray}
{\Gamma _{b{\rm{k}}}}\left( t \right) &=& + {\Gamma _{b{\rm{k}}}}\left( {{t_0}} \right){e^{ - i{\omega _{b{\rm{k}}}}\left( {t - {t_0}} \right)}} \nonumber\\&&- i{g_{b{\rm{k}}}}\int_{{t_0}}^t {d\tau } \left[ {b\left( \tau  \right) - {\beta _{\rm{s}}}} \right]{e^{ - i{\omega _{b{\rm{k}}}}\left( {t - \tau } \right)}}. \label{eq-106}
\end{eqnarray}
Here the first term describes the free evolution of the reservoir modes, whereas the second term arises from their interaction with the mechanical mode. We eliminate the reservoir operators ${\Gamma _{b{\rm{k}}}}$ by substituting Eq.~(\ref{eq-106}) into Eq.~(\ref{eq-104}), finding
\begin{eqnarray}
\dot b &=&  - i{\Delta _{\rm{m}}}b + ig_{\rm{s}}{a^\dag }a - i\left( {{\Omega _b} - \beta_{\rm{s}} {\Delta _{\rm{m}}}} \right) \nonumber\\
 &&- \sum\limits_{\rm{k}} {g_{b{\rm{k}}}^2} \int_{{t_0}}^t d \tau  \left[b\left( \tau  \right) - \beta_{\rm{s}} \right]{e^{ - i{\omega _{b{\rm{k}}}}\left( {t - \tau } \right)}} \nonumber\\ &&- i\sum\limits_{\rm{k}} {{g_{b{\rm{k}}}}} {\Gamma _{b{\rm{k}}}}\left( {{t_0}} \right){e^{ - i{\omega _{b{\rm{k}}}}\left( {t - {t_0}} \right)}}. \label{eq-107}
\end{eqnarray}
In Eq.~(\ref{eq-107}), we can see that the evolution of the system operator depends on the fluctuations in the reservoir. Next, we make some approximations. As in the Weisskopf-Wigner approximation \cite{ref-58}, we consider the spectrum to be given by the normal modes of a large scale, and these modes are very close in frequency. We then approximate this spectrum by a continuous spectrum. Thus, the summation in Eq.~(\ref{eq-107}) can be written as
\begin{eqnarray}
\dot b &=&  - i{\Delta _{\rm{m}}} + i{g_{\rm{s}}}{a^\dag }a - i\left( {{\Omega _b} - {\beta _{\rm{s}}}{\Delta _{\rm{m}}}} \right) \nonumber\\ &&- i\int_0^{ + \infty } \!\!\!\!\!\!\! {{g_b}} \left( {{\omega _b}} \right)\frac{{d{\rm{k}}\left( {{\omega _b}} \right)}}{{d{\omega _b}}}{\Gamma _b}\left( {{\omega _b},{t_0}} \right){e^{ - i{\omega _b}\left( {t - {t_0}} \right)}}d{\omega _b} \label{eq-108}\\
 &&- \int_{{t_0}}^t \!{\int_0^{ + \infty } \!\!\!\!\!\!\!\! {g_b^2} } \left( {{\omega _b}} \right)\frac{{d{\rm{k}}\left( {{\omega _b}} \right)}}{{d{\omega _b}}}\left[ {b\left( \tau  \right) - {\beta _{\rm{s}}}} \right]{e^{ - i{\omega _b}\left( {t - {\tau}} \right)}}d{\omega _b}d\tau. \nonumber
\end{eqnarray}
Considering an ideal situation, we assume for simplicity that ${{{{\left[ {{g_b}\left( {{\omega _b}} \right)} \right]}^2}d{\rm{k}}\left( {{\omega _b}} \right)} \mathord{/{\vphantom {{{{\left[ {{g_b}\left( {{\omega _b}} \right)} \right]}^2}d{\rm{k}}\left( {{\omega _b}} \right)} d}} \kern-\nulldelimiterspace} d}{\omega _b} \!\!=\! {{{\gamma _b}} \mathord{\left/{\vphantom {{{\gamma _b}} {2\pi }}} \right.\kern-\nulldelimiterspace} {2\pi }} \!>\!\! 0$ is constant, so that Eq.~(\ref{eq-108}) is reduced to a simple first-order differential equation \cite{ref-59}:
\begin{eqnarray}
\dot b &=&  - i{\Delta _{\rm{m}}} + i{g_{\rm{s}}}{a^\dag }a - i\left( {{\Omega _b} - {\beta _{\rm{s}}}{\Delta _{\rm{m}}}} \right) \nonumber\\ &&- i\sqrt {\frac{{{\gamma _b}}}{{2\pi }}} \int_0^{ + \infty } {{\Gamma _b}\left( {{\omega _b},{t_0}} \right){e^{ - i{\omega _b}\left( {t - {t_0}} \right)}}d{\omega _b}} \nonumber\\
 &&- \frac{{{\gamma _b}}}{{2\pi }}\int_{{t_0}}^t {\int_0^{ + \infty } {\left[ {b\left( \tau  \right) - {\beta _{\rm{s}}}} \right]{e^{ - i{\omega _b}\left( {t - {t_0}} \right)}}} } d{\omega _b}d\tau. \label{eq-109}
\end{eqnarray}
Using the relation
\begin{eqnarray}
\int_{ 0 }^{ + \infty } {d{\omega _b}} {e^{ - i{\omega _b}\left( {t - \tau } \right)}} = \pi \delta \left( {t - \tau } \right),\label{eq-110}
\end{eqnarray}
we arrive at Eq.~(\ref{eq-8}) in the main text:
\begin{eqnarray}
\dot b =  - \left( {i{\Delta _{\rm{m}}} + \frac{{{\gamma _b}}}{2}} \right)b + ig{a^\dag }a + \sqrt {{\gamma _b}} {b_{{\rm{in}}}} + \dot \beta_{\rm{s}}\label{eq-110}
\end{eqnarray}
with
\begin{eqnarray}
{b_{{\rm{in}}}}\left( t \right) &=&  - \frac{i}{{\sqrt {2\pi } }}\int_0^{ + \infty } {{\Gamma _b}\left( {{\omega _b},{t_0}} \right){e^{ - i{\omega _b}\left( {t - {t_0}} \right)}}d{\omega _b}}, \nonumber\\ \dot \beta_{\rm{s}}  &=& \frac{{{\gamma _b}}}{2}\beta_{\rm{s}}  - i\left( {{\Omega _b} - \beta_{\rm{s}} {\Delta _{\rm{m}}}} \right) = 0, \label{eq-111}
\end{eqnarray}
where ${b_{{\rm{in}}}}$ is a noise operator which depends upon the initial-time environment operators ${\Gamma _b}\left( {{\omega _b},{t_0}} \right)$, while ${\gamma _b}$ is the mechanical damping rate which depends on the coupling strength $g_{b{\rm{k}}}$ of the system and the reservoir. This produces Eq.~(\ref{eq-9}) in the main text. The same holds true for Eq.~(\ref{eq-8}).

\section{An alternative viewpoint on the selection of ${\beta _{\rm{s}}}$} \label{appendix C}
In this Appendix, we present the physical meaning of the coherent displacement amplitude $\beta $ at a steady state from the point of view of the Gorini-Kossakowski-Sudarshan-Lindblad (GKSL) equation \cite{ref-57,ref-58}. The derivation here is based on a recent related work \cite{ref-56}. To begin with, let us consider an optical mode $a$ and a mechanical mode $b$ which are coupled to two individual Markovian heat baths consisting of assemblies of the oscillators. When the reservoir modes ${\Gamma _{a{\rm{k}}}}$ and ${\Gamma _{b{\rm{k}}}}$ are initially in the thermal equilibrium, in the presence of dissipations, the evolution of the field system ${\rho _{\rm{S}}}$ is governed by the GKSL master equation as follows:
\begin{eqnarray}
{{\dot \rho }_{\rm{S}}} &=& \frac{i}{\hbar }\left[ {{\rho _{\rm{S}}},H_{\rm{S}}^{\rm{O}}} \right]{\rm{ + }}{\kappa _a}\left( {{{\bar n}_a} + 1} \right)L\left[ a \right]{\rho _{\rm{S}}} + {\kappa _a}{{\bar n}_a}L\left[ {{a^\dag }} \right]{\rho _{\rm{S}}} \nonumber\\&&+ {\gamma _b}\left( {{{\bar n}_b} + 1} \right)L\left[ b \right]{\rho _{\rm{S}}} + {\gamma _b}{{\bar n}_b}L\left[ {{b^\dag }} \right]{\rho _{\rm{S}}}, \label{eq-113}
\end{eqnarray}
where the standard Lindblad superoperator for the optical mode damping can be written as $L\left[ a \right]{\rho _{\rm{S}}} = a{\rho _{\rm{S}}}a - ( {{a^\dag }a{\rho _{\rm{S}}} + {\rho _{\rm{S}}}{a^\dag }a} )$, and the same goes for the mechanical mode damping. Here, ${\rho _{\rm{S}}}$ denotes the reduced density operator for the open system (\ref{eq-3}), ${\bar n_a}$ and ${\bar n_b}$ denote the number operators of the reservoirs corresponding to the optical mode $a$ and mechanical mode $b$, respectively, ${\kappa _a}$ denotes the decay rate of the optical cavity and ${{\gamma _b}}$ denotes the mechanical damping rate.

When a high-power laser drives the mechanical mode, we can make the following displacement transformation in the coherent-state representation:
\begin{eqnarray}
{\rho '_{\rm{S}}} = D\left( \beta  \right){\rho _{\rm{S}}}{D^\dag }\left( \beta  \right), \label{eq-114}
\end{eqnarray}
where ${\rho '_{\rm{S}}}$ is the reduced density operator for the closed system ${H_{\rm{S}}^{\rm{P}}}$ in the displacement representation. In the displacement operator $D\left( \beta  \right) = \exp \left( {\beta {b^\dag } - {\beta ^*}b} \right)$, $\beta $ is the amplitude of the coherent displacement, which needs to be determined in the transformed master equation.

We now try to find the GKSL master equation for ${\rho '_{\rm{S}}}$. The time derivative of Eq.~(\ref{eq-114}) reads
\begin{eqnarray}
\!\!{{\dot \rho '}_{\rm{S}}} \!=\! \dot D\left( \beta  \right)\!{\rho _{\rm{S}}}{D^\dag }\left( \beta  \right) \!+\! D\left( \beta  \right)\!{{\dot \rho }_{\rm{S}}}{D^\dag }\left( \beta  \right) \!+\! D\left( \beta  \right)\!{\rho _{\rm{S}}}{{\dot D}^\dag }\!\!\left( \beta  \right) \!. \label{eq-115}
\end{eqnarray}
Using the relations $D\left( \beta  \right){D^\dag }\left( \beta  \right) = {D^\dag }\left( \beta  \right)D\left( \beta  \right) = \mathbb{I}$ and $D\left( \beta  \right) = \exp \left( {{{\beta {\beta ^*}} \mathord{\left/{\vphantom {{\beta {\beta ^*}} 2}} \right.\kern-\nulldelimiterspace} 2}} \right)\exp \left( { - b{\beta ^*}} \right)\exp \left( {{b^\dag }\beta } \right)$, we find \!$\dot D\left( \beta  \right)$ and ${{\dot D}^\dag }\!\!\left( \beta  \right)$ as in
\begin{eqnarray}
\dot D\left( \beta  \right) \!\!&=&\!\! \frac{1}{2}\left( {\dot \beta {\beta ^*} - \beta {{\dot \beta }^*}} \right)\!D\left( \beta  \right) \!+\! D\!\left( \beta  \right)\left( {\dot \beta {b^\dag } - {{\dot \beta }^*}b} \right)\!\!, \label{eq-116}\\
{{\dot D}^\dag }\left( \beta  \right) \!\!&=&\!\!  - \frac{1}{2}\left( {\dot \beta {\beta ^*} - \beta {{\dot \beta }^*}} \right)\!{D^\dag }\left( \beta  \right) \!+\! \left( {{{\dot \beta }^*}b - \dot \beta {b^\dag }} \right)\!{D^\dag }\left( \beta  \right). \nonumber
\end{eqnarray}
We can thereby reduce Eq.~(\ref{eq-115}) to
\begin{eqnarray}
\!\! {{\dot \rho '}_{\rm{S}}} = D\left( \beta  \right){{\dot \rho }_{\rm{S}}}{D^\dag }\left( \beta  \right) \!-\! D\!\left( \beta  \right)\!\left[ {{\rho _{\rm{S}}},\left( {\dot \beta {b^\dag } - {{\dot \beta }^*}b} \right)} \right]\!{D^\dag }\!\!\left( \beta  \right)\!\!. \label{eq-117}
\end{eqnarray}
Then, by using the expressions $D\left( \beta  \right)b{D^\dag }\left( \beta  \right) = b - \beta $ and $D\left( \beta  \right){b^\dag }{D^\dag }\left( \beta  \right) \!=\! {b^\dag } - {\beta ^*}$ and inserting Eq.~(\ref{eq-113}) into Eq.~(\ref{eq-117}), we find the transformed GKSL master equation in the form
\begin{eqnarray}
{\dot \rho '_{\rm{S}}} &=& \frac{i}{\hbar }\left[ {{\rho '_{\rm{S}}},H_{\rm{S}}^{{\rm{O'}}}} \right]{\rm{ + }}{\kappa _a}\left( {{{\bar n}_a} + 1} \right)L\left[ a \right]{\rho '_{\rm{S}}} + {\kappa _a}{{\bar n}_a}L\left[ {{a^\dag }} \right]{\rho '_{\rm{S}}} \nonumber\\&&+ {\gamma _b}\left( {{{\bar n}_{\rm{m}}} + 1} \right)L\left[ b \right]{\rho '_{\rm{S}}} + {\gamma _b}{{\bar n}_{\rm{m}}}L\left[ {{b^\dag }} \right]{\rho '_{\rm{S}}}
\nonumber\\&&+ \left[ {\dot \beta  + \left( {i{\Delta _{\rm{m}}} + \frac{{{\gamma _b}}}{2}} \right)\beta  - i{\Omega _b}} \right]\left[ {{b^\dag },{\rho '_{\rm{S}}}} \right] \label{eq-118}\\&&- \left[ {{{\dot \beta }^*} + \left( { - i{\Delta _{\rm{m}}} + \frac{{{\gamma _b}}}{2}} \right){\beta ^*} + i\Omega _b^*} \right]\left[ {b,{\rho '_{\rm{S}}}} \right], \nonumber
\end{eqnarray}
where
\begin{eqnarray}
H_{\rm{S}}^{{\rm{O'}}} &=& +\hbar \left[ {{\Delta _{\rm{c}}} + \chi {{\left| \beta  \right|}^2} + 2{g_0}{\mathop{\rm Re}\nolimits} \left( \beta  \right)} \right]{a^\dag }a \nonumber\\&&+ \hbar \left( {{\Delta _{\rm{m}}} + \chi {a^\dag }a} \right){b^\dag }b  - \hbar \chi {a^\dag }a\left( {\beta {b^\dag } + {\beta ^*}b} \right) \nonumber\\&&- \hbar {g_0}{a^\dag }a\left( {{b^\dag } + b} \right) + \hbar \Omega _a^*a + \hbar {\Omega _a}{a^\dag }.\label{eq-119}
\end{eqnarray}
We can see that the condition for the GKSL master equation to remain unchanged after the coherent displacement transformation is
\begin{eqnarray}
\dot \beta \left( t \right) + \left( {i{\Delta _{\rm{m}}} + \frac{{{\gamma _b}}}{2}} \right)\beta \left( t \right) - i{\Omega _b} = 0. \label{eq-120}
\end{eqnarray}
By solving Eq.~(\ref{eq-120}), we obtain the solution of the amplitude of the transient displacement as follows:
\begin{eqnarray}
\beta \left( t \right) = \frac{{{\Omega _b}}}{{{\Delta _{\rm{m}}} - i0.5{\gamma _b}}} + C\left( {{t_0}} \right){e^{ - \left( {i{\Delta _{\rm{m}}} + 0.5{\gamma _b}} \right)t}},\label{eq-121}
\end{eqnarray}
In the limit $t \to  + \infty $, the amplitude is reduced to the steady-state solution
\begin{eqnarray}
{\beta _{\rm{s}}} = \frac{{{\Omega _b}}}{{{\Delta _{\rm{m}}} - i0.5{\gamma _b}}}.\label{eq-122}
\end{eqnarray}
We can see from Eq.~(\ref{eq-122}) that the amplitude of steady-state coherent displacement ${\beta _{\rm{s}}}$ is tunable by choosing proper parameters ${{\Omega _b}}$ and ${{\Delta _{\rm{m}}}}$. Next, taking ${\beta _{\rm{s}}}$, we obtain the GKSL master equation in the displacement representation as in
\begin{eqnarray}
{{\dot \rho '}_{\rm{S}}} &=& \frac{i}{\hbar }\left[ {{\rho '_{\rm{S}}},H_{\rm{S}}^{{\rm{O''}}}} \right]{\rm{ + }}{\kappa _a}\left( {{{\bar n}_a} + 1} \right)L\left[ a \right]{\rho '_{\rm{S}}} + {\kappa _a}{{\bar n}_a}L\left[ {{a^\dag }} \right]{\rho '_{\rm{S}}} \nonumber\\ &&+ {\gamma _b}\left( {{{\bar n}_{\rm{m}}} + 1} \right)L\left[ b \right]{\rho '_{\rm{S}}} + {\gamma _b}{{\bar n}_{\rm{m}}}L\left[ {{b^\dag }} \right]{\rho '_{\rm{S}}},\label{eq-123}
\end{eqnarray}
for which the steady-state transformed Hamiltonian in the displacement representation becomes
\begin{eqnarray}
H_{\rm{S}}^{{\rm{O''}}} &=& +\hbar \Delta _{\rm{c}}^0{a^\dag }a + \hbar \left( {{\Delta _{\rm{m}}} + \chi {a^\dag }a} \right){b^\dag }b \nonumber\\&&- \hbar {a^\dag }a\left( {{g_{\rm{s}}}{b^\dag } + g_{\rm{s}}^*b} \right) + \hbar \Omega _a^*a + \hbar {\Omega _a}{a^\dag }\label{eq-124}
\end{eqnarray}
with $\Delta _{\rm{c}}^0 = {\Delta _{\rm{c}}} + \chi {\left| {{\beta _{\rm{s}}}} \right|^2}$ being the normalized optical cavity detuning including the frequency shift caused by the high-power laser driving and ${g_{\rm{s}}} = \chi {\beta _{\rm{s}}}$. Since our motivation is to study the few-photon physics in strong optomechanical system, we demand the constraint condition $\chi  \approx {10^{ - 3}}{\Delta _{\rm{m}}}$ in circuit-QED \cite{ref-50,ref-52,ref-56}. Thus, we neglect the enhanced CK interaction term in Eq.~(\ref{eq-124}) to obtain the approximate Hamiltonian, $H_{\rm{S}}^{{\rm{app}}}$ as shown in the main text.

\section{Scattering Probability modification for non-RWA case with ultratrong coupling} \label{appendix D}

In this Appendix, we prove numerically that Eq.~(\ref{eq-46}) is more universal than the results in Ref.~\cite{ref-81} for non-RWA cases after a slight modification. The expression of the scattering probability in Ref.~\cite{ref-81} is only successful in the cases of weak coupling and strong coupling. However, in the case of ultrastrong coupling, it needs to be modified to Eq.~(\ref{eq-46}).

The numerical validation process is as follows. In the non-RWA case, the scattering probability from $a$ to $b$ described in Ref.~\cite{ref-81} is
\begin{eqnarray}
\sigma _a^b\left( \omega  \right) = |{\rm{O}}_a^b\left( \omega  \right){|^2} + |{\rm{O}}_a^{{b^\dag }}\left( \omega  \right){|^2}.\label{eq-125}
\end{eqnarray}

\begin{figure}[b]
\centering
\includegraphics[angle=0,width=0.46\textwidth]{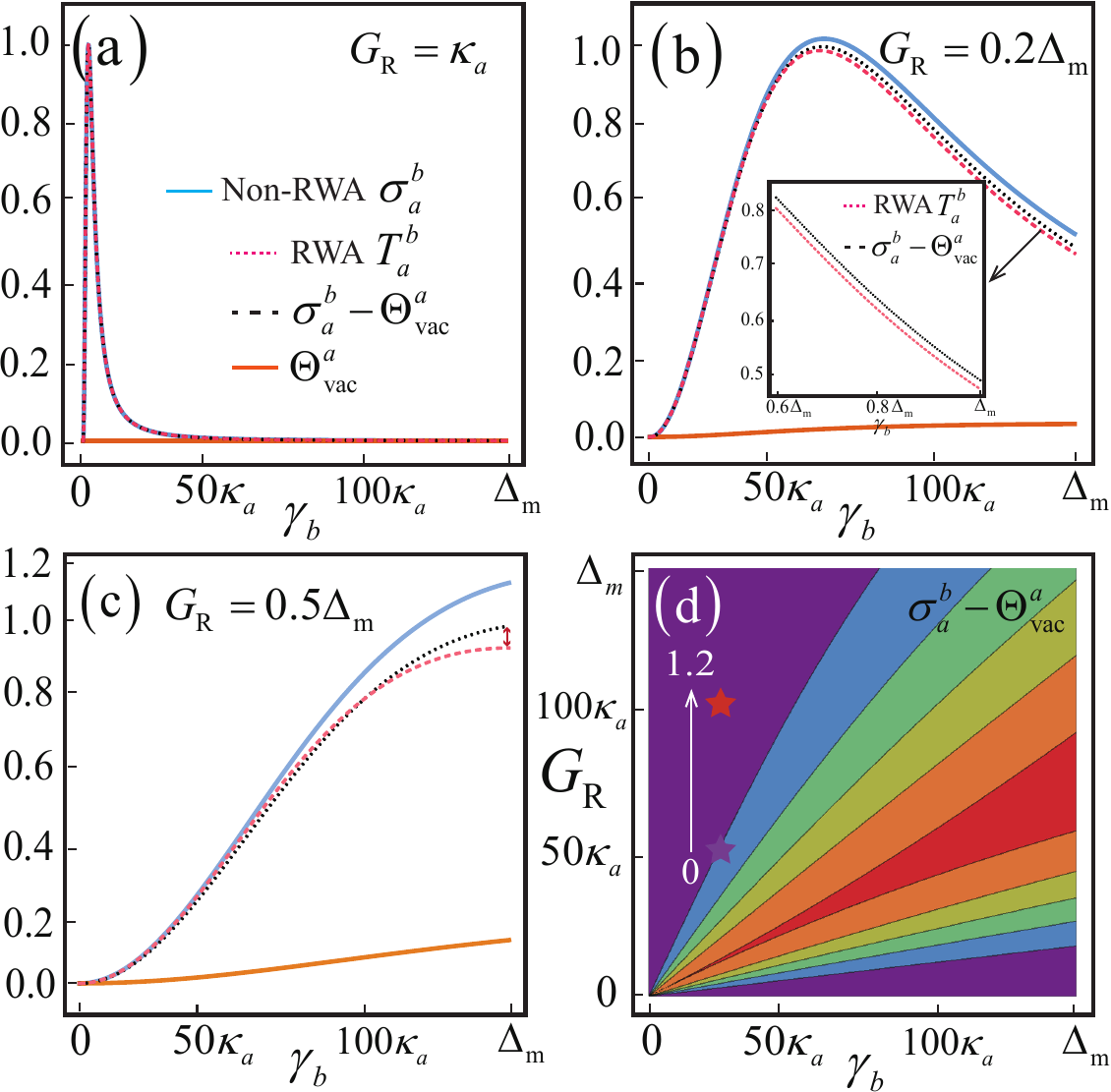}
\caption{We compare the scattering probabilities $\sigma _a^b$ before the RWA in Eq.~(\ref{eq-125}) and $T_a^b$ after it as functions of the mechanical damping ${{\gamma _b}}$ for different effective optomechanical coupling ${G_R}$: $\left( \rm{a} \right)$ ${G_R} = {\kappa _a}$, $\left( \rm{b} \right)$ ${G_R} = 0.2{\Delta _{\rm{m}}}$, $\left( \rm{c} \right)$ ${G_R} = 0.5{\Delta _{\rm{m}}}$. We show the scattering probability $\sigma _a^b - \Theta _{{\rm{vac}}}^a$ as a function of ${G_R}$ and ${\gamma _b}$ in $\left( {\rm{d}} \right)$. The other parameters are set to ${\kappa _a} = {\gamma _b} \in \left( {0,{\Delta _{\rm{m}}}} \right)$ and $\omega  = {\Delta ''_c} = {\Delta _{\rm{m}}}$.} \label{Fig-13}
\end{figure}

Using the same values of parameters as in Fig.~\ref{Fig-10}$\left( {\rm{a}} \right)$-$\left( {\rm{c}} \right)$, we use Eq.~(\ref{eq-125}) to numerically calculate the scattering probability of input signal from weak coupling to ultrastrong coupling cases in Fig.~\ref{Fig-13}$\left( {\rm{a}} \right)$-$\left( {\rm{c}} \right)$. When ${G_R} = {\kappa _a}$, the equality $\sigma _a^b - \Theta _{{\rm{vac}}}^a = T_a^b$ is always true, which is consistent with the cases of weak coupling and strong coupling discussed in Ref.~\cite{ref-81}. However, in the ultrastrong coupling case, when ${G_R} = 0.2{\Delta _{\rm{m}}}$ and ${G_R} = 0.5{\Delta _{\rm{m}}}$, under strong dissipation, ${\rm{max}}\left( {\sigma _a^b - \Theta _{{\rm{vac}}}^a} \right) > {\rm{max}}\left( {T_a^b} \right) = 1$, which violates the law of conservation of energy \cite{ref-98,ref-99}. Compared with it, Eq.~(\ref{eq-46}) satisfies ${\rm{P}}_a^b - \Theta _{{\rm{vac}}}^a = T_a^b$ under any coupling strength; see Fig.~\ref{Fig-10}.


\begin{references}
\bibitem{ref-1} M. Aspelmeyer, T. J. Kippenberg, and F. Marquardt, \textit{Cavity Optomechanics: Nano- and Micromechanical Resonators Interacting with Light} (Springer, Germany, 2014).
\bibitem{ref-2} M. Aspelmeyer, T. J. Kippenberg, and F. Marquardt, Cavity optomechanics, Rev. Mod. Phys. \textbf{86}, 1391 (2014).
\bibitem{ref-3} W. P. Bowen and G. J. Milburn, \textit{Quantum optomechanics} (CRC, USA, 2016).
\bibitem{ref-4} T. J. Kippenberg and K. J. Vahala, Cavity optomechanics: Back-action at the mesoscale, Science \textbf{321}, 1172 (2008).
\bibitem{ref-5} F. Marquardt and S. M. Girvin, Optomechanics, Physics \textbf{2}, 40 (2009).
\bibitem{ref-6} M. Aspelmeyer, P. Meystre, and K. Schwab, Quantum optomechanics, Phys. Today \textbf{65}(7), 29 (2012).
\bibitem{ref-7} A. Nunnenkamp, K. B{\o}rkje, and S. M. Girvin, Single-Photon Optomechanics, Phys. Rev. Lett. \textbf{107}, 063602 (2011).
\bibitem{ref-8} T. Hong, H. Miao, and Y. Chen, Open quantum dynamics of single-photon optomechanical devices, Phys. Rev. A \textbf{88}, 023812 (2013).
\bibitem{ref-9} H. X. Tang and D. Vitali, Prospect of detecting single-photon-force effects in cavity optomechanics, Phys. Rev. A \textbf{89}, 063821 (2014).
\bibitem{ref-10} J.-Q. Liao, H. K. Cheung, and C. K. Law, Spectrum of single-photon emission and scattering in cavity optomechanics, Phys. Rev. A \textbf{85}, 025803 (2012).
\bibitem{ref-11} P. Rabl, Photon Blockade Effect in Optomechanical Systems, Phys. Rev. Lett. \textbf{107}, 063601 (2011).
\bibitem{ref-12} J.-Q. Liao and F. Nori, Photon blockade in quadratically coupled optomechanical devices, Phys. Rev. A \textbf{88}, 023853 (2013).
\bibitem{ref-13} H. Z. Shen, Cheng Shang, Y. H. Zhou, and X. X. Yi, Unconventional single-photon blockade in non-Markovian systems, Phys. Rev. A \textbf{98}, 023856 (2018).
\bibitem{ref-14} H. Y. Sun, Cheng Shang, X. X. Luo, Y. H. Zhou, and H. Z. Shen, Optical-assisted Photon Blockade in a Cavity System via Parametric Interactions, Int. J. Theor. Phys \textbf{58}, 3640 (2019).
\bibitem{ref-15} J.-Q. Liao and L. Tian, Macroscopic Quantum Superposition in Cavity Optomechanics, Phys. Rev. Lett. \textbf{116}, 163602 (2016).
\bibitem{ref-16} W. Marshall, C. Simon, R. Penrose, and D. Bouwmeester, Towards Quantum Superpositions of a Mirror, Phys. Rev. Lett. \textbf{91}, 130401 (2003).
\bibitem{ref-17} Max Ludwig, B. Kubala, and F. Marquardt, The optomechanical instability in the quantum regime, New J. Phys. \textbf{10}, 095013 (2008).
\bibitem{ref-18} F. Brennecke, S. Ritter, T. Donner, and T. Esslinger, Cavity Optomechanics with a Bose-Einstein Condensate, Science \textbf{322}, 235 (2008).
\bibitem{ref-19} A. Xuereb, C. Genes, and A. Dantan, Strong Coupling and Long-Range Collective Interactions in Optomechanical Arrays, Phys. Rev. Lett. \textbf{109}, 223601 (2012).
\bibitem{ref-20} T. P. Purdy, P.-L. Yu, R. W. Peterson, N. S. KaMPEL, and C. A. Regal, Strong Optomechanical Squeezing of Light, Phys. Rev. X \textbf{3}, 031012 (2013).
\bibitem{ref-21} X.-Y L\"{u}, Y. Wu, J. R. Johansson, H. Jing, J. Zhang, and F. Nori, Squeezed Optomechanics with Phase-Matched Amplification and Dissipation, Phys. Rev. Lett. \textbf{114}, 093602 (2015).
\bibitem{ref-22} M.-A. Lemonde, N. Didier, and A. A. Clerk, Enhanced non-linear interactions in quantum optomechanics via mechanical amplification, Nat. Commun. \textbf{7}, 11338 (2016).
\bibitem{ref-23} Z. Y. Wang and A. H. S.-Naeini, Enhancing a slow and weak optomechanical nonlinearity with delayed quantum feedback, Nat. Commun. \textbf{8}, 15886 (2017).
\bibitem{ref-24} W. Xiong, J. J. Chen, B. L. Fang, M. F. Wang, L. Ye, and J. Q. You, Strong tunable spin-spin interaction in a weakly coupled nitrogen vacancy spin-cavity electromechanical system, Phys. Rev. B \textbf{103}, 174106 (2021).
\bibitem{ref-25} X.-L. Yin, Y.-H. Zhou, and J.-Q. Liao, All-optical quantum simulation of ultrastrong optomechanics, Phys. Rev. A \textbf{105}, 013504 (2022).
\bibitem{ref-26} I. S\"{o}llner, L. Midolo, and P. Lodahl, Deterministic Single-Phonon Source Tiggered by a Single Photon, Phys. Rev. Lett. \textbf{116}, 234301 (2016).
\bibitem{ref-27} R. Khan, F. Massel, T. T. Heikkil\"{a}, Cross-Kerr nonlinearity in optomechanical systems, Phys. Rev. A \textbf{91}, 043822 (2015).
\bibitem{ref-28} S. Ding, G. Maslennikov, R. Habl\"{u}tze, and D. Matsukevich, Cross-Kerr Nonlinearity for Phonon Counting, Phys. Rev. Lett. \textbf{119}, 193602 (2017).
\bibitem{ref-29} C. K. Law, Effective Hamiltonian for the radiation in a cavity with a moving mirror and a time-varying dielectric medium, Phys. Rev. A \textbf{49}, 433 (1994).
\bibitem{ref-30} R. Sarala, F. Massel, Cross-Kerr nonlinearity: a stability analysis, Nanoscale Systems: Mathematical Modeling, Theory and Application, \textbf{4}(1), 2299 (2015).
\bibitem{ref-31} F. Zhou, L. B. Fan, J.-F. Huang, and J.-Q. Liao, Enhancement of few-photon optomechanical effects with cross-Kerr nonlinearity, Phys. Rev. A \textbf{99}, 043837 (2019).
\bibitem{ref-32} M. A. Lemonde, N. Didier, and A. A. Clerk, Nonlinear Interaction Effects in a Strongly Driven Optomechanical Cavity, Phys. Rev. Lett. \textbf{111}, 053602 (2013).
\bibitem{ref-33} J.-S. Zhang, M.-C. Li, and A.-X. Chen, Enhancing quadratic optomechanical coupling via a nonlinear medium and lasers, Phys. Rev. A \textbf{99}, 013843 (2019).
\bibitem{ref-34} S. Chakraborty and A. K. Sarma, Enhancing quantum correlations in an optomechanical system via cross-Kerr nonlinearity, J. Opt. Soc. Am. B \textbf{34}, 1503 (2017).
\bibitem{ref-35} P. F.-D\'{i}az, L. Lammata, E. Rico, and E. Solano, Ultrastrong coupling regimes of light-matter interaction, Rev. Mod. Phys. \textbf{91}, 025005 (2019).
\bibitem{ref-36} A. F. Kockum, A. Miranowicz, S. D. Liberato, S. Savasta, and F. Nori, Ultrastrong coupling between light and matter, Nature Reviews Physics \textbf{1}, 19 (2019).
\bibitem{ref-37} J. R. Carson, A generalization of reciprocal theorem, Bell Sys. Tech. J. \textbf{3}, 393 (1924).
\bibitem{ref-38} R. Carminati, J. J. S\'{a}enz, J.-J. Greffet, and M. N.-Vesperinas, Reciprocity, unitarity, and time-reversal symmetry of the \textit{S} matrix of fields containing evanescent components, Phys. Rev. A \textbf{62}, 012712 (2000).
\bibitem{ref-39} M. N.-Vesperinas, \textit{Scattering and Diffraction in Physical Optics} (Word Scientific, Singapore, 2006).
\bibitem{ref-40} A. A. Anappara, S. D. Liberato, Alessandro Tredicucci, Cristiano Ciuti, Giorgio Biasiol, Lucia Sorba, and Fabio Beltram, Signatures of the ultrastrong light-matter coupling regime. Phys. Rev. B \textbf{79}, 201303 (2009).
\bibitem{ref-41} C. K. Law, Interaction between a movinng mirror and radition pressure: A Hamiltonian formulation, Phys. Rev. A \textbf{51}, 2537 (1995).
\bibitem{ref-42} A. Schliesser, R. Rivi\`{e}re, G. Anetsberger, O. Arcizet, and T. J. Kippenberg, Resolved-sideband cooling of a micromechanical oscillator, Nature Physics \textbf{4}, 415 (2008).
\bibitem{ref-43} K. Koshino, Semiclassical evaluation of the two-photon cross-Kerr effect, Phys. Rev. A \textbf{74}, 053818 (2006).
\bibitem{ref-44} L. Mandel and E. Wolf, \textit{Optical Coherence and Quantum Optics} (Cambridge University, UK, 1995).
\bibitem{ref-79} G. S. Agarwal, and S. Huang, Optomechanical systems as single-photon routers, Phys. Rev. A \textbf{85}, 021801 (R) (2012).
\bibitem{ref-80} Darrick E. Chang, Anders S. S{\o}rensen, Eugene A. Demler, and Mikhail D. Lukin, A single-photon transistor using nanoscale surface plasmons, Nature Physics \textbf{3}, 807-812 (2007).
\bibitem{ref-45} T. T. Heikkil\"{a}, F. Massel, J. Tuorila, R. Khan, and M. A. Sillanp\"{a}\"{a}, Enhancing Optomechanical Coupling via the Josephson Effect, Phys. Rev. Lett. \textbf{112}, 203603 (2014).
\bibitem{ref-46} W. Xiong, D.-Y. Jin, Y. Qiu, C.-H. Lam, and J. Q. You, Cross-Kerr effect on an optomechanical system, Phys. Rev. A \textbf{93}, 023844 (2016).
\bibitem{ref-88} Z. L. Xiang, S. Ashhab, J. Q. You, and F. Nori, Hybrid quantum circuits: Superconducting circuits interacting with other quantum systems, Rev. Mod. Phys. \textbf{85}, 623 (2013).
\bibitem{ref-89} J. Q. You and F. Nori, Atomic physics and quantum optics using superconducting circuits, Nature(London) \textbf{474}, 589 (2011).
\bibitem{ref-90} J. M. Pirkkalainen, S. U. Cho, F. Massel, J. Tuorila, T. T. Heikkil\"{a}, P . J. Hakonen, Cavity optomechanics mediated by a quantum two-level system,  Nat. Commun. \textbf{6}, 6981 (2015).
\bibitem{ref-91} M. A. Sillanp\"{a}\"{a}, Leif Roschier, and P. J. Hakonen, Inductive single-electron transistor (L-SET). Phys. Rev. Lett. \textbf{93}, 066805 (2004).
\bibitem{ref-47} Dustin Kleckner, William Marshall, Michiel J. A. de Dood, Khodadad Nima Dinyari, Bart-Jan Pors, William T. M. Irvine, and Dirk Bouwmeester, High Finesse Opto-Mechanical Cavity with a Movable Thirty-Micron-Size Mirror, Phys. Rev. Lett. \textbf{96}, 173901 (2006).
\bibitem{ref-48} D. Vitali, S. Gigan, A. Ferreira, H. R. B\"{o}hm, P. Tombesi, A. Guerreiro, V. Vedral, A. Zeilinger, and M. Aspelmyer, Optomechanical Entanglement between a Movable Mirror and a Cavity Field, Phys. Rev. Lett. \textbf{98}, 030405 (2007).
\bibitem{ref-49} Y. Hu, G.-Q. Ge, S. Chen, X.-F. Yang, and Y.-L. Chen, Cross-Kerr-effect induced by coupled Josephson qubits in circuit quantum electrodynamics, Phys. Rev. A \textbf{84}, 012329 (2011).
\bibitem{ref-Lu053703} Zhi-Guang Lu, Cheng Shang, Ying Wu, and Xin-You Lü, Analytical approach to higher-order correlation functions in U(1) symmetric systems, Phys. Rev. A 108, 053703 (2023).
\bibitem{ref-50} E. T. Holland, B. Vlastakis, R. W. Heeres, M. J. Reagor, U. Vool, Z. Leghtas, L. Frunzio, G. Kirchmair, M. H. Devoret, M. Mirrahimi, and R. J. Schoelkopf, Single-Photon-Resolved Cross-Kerr Interaction for Autonomous Stabilization of Photon-Number States, Phys. Rev. Lett. \textbf{115}, 180501 (2015).
\bibitem{ref-52} J. Majer, J. M. Chow, J. M. Gambetta, Jens Koch, B. R. Johnson, J. A. Schreier, L. Frunzio, D. I. Schuster, A. A. Houck, A. Wallraff, A. Blais, M. H. Devoret, S. M. Girvin, and R. J. Schoelkopf, Coupling superconducting qubits via a cavity bus, Nature (London) \textbf{449}, 443 (2007).
\bibitem{ref-53} D. F. Walls and G. J. Milbrum, \textit{Quantum Optics} (Springer, Germany, 2008).
\bibitem{ref-54} S. M. Barnett and P. M. Radmore, \textit{Methods in Theoretical Quantum Optics} (Clarendon Press, Oxford, 1997).
\bibitem{ref-55} S. Gr\"{o}blacher, K. Hammerer, M. R. Vanner, and M. Aspelmeyer, Observation of strong coupling between a micromechanical resonator and an optical cavity field, Nature (London) \textbf{460}, 724 (2009).
\bibitem{ref-56} J.-Q. Liao, J.-F. Huang, L. Tian, L.-M. Kuang, and C. P. Sun, Generalized ultrastrong optomechanical-like coupling, Phys. Rev. A \textbf{101}, 063802 (2020).
\bibitem{ref-57} A. A. Clerk, M. H. Devoret, S. M. Girvin, F. Marquardt, and R. J. Schoelkopf, Introduction to quantum noise, measurement, and amplification, Rev. Mod. Phys. \textbf{82}, 1155 (2010).
\bibitem{ref-58} M. O. Scully and M. S. Zubairy, \textit{Quantum Optics} (Cambridge University, UK, 2011).
\bibitem{ref-59} C. W. Gardiner and P. Zoller, \textit{Quantum noise}, Germany (2000).
\bibitem{ref-60} Daniel Manzano, A short introduction to the Lindblad master equation, AIP Advances \textbf{10}, 025106 (2020).
\bibitem{ref-Lu180401} Zhi-Guang Lu, Guoqing Tian, Xin-You Lü, and Cheng Shang, Topological Quantum Batteries, Phys. Rev. Lett. \textbf{134}, 180401 (2025).
\bibitem{ref-Li250313731} Hongchao Li, Cheng Shang, Tomotaka Kuwahara, and Tan Van Vu, Macroscopic Particle Transport in Dissipative Long-Range Bosonic Systems, arXiv: 2503.13731 (2025).
\bibitem{ref-61} V. Gorini, A. Kossakowski, and E. C. Sudarsahan, Completely positive semigroups of n-level systems, J. Math. Phys. \textbf{17}, 821 (1976).
\bibitem{ref-62} G. M. Moy, J. J. Hope, and C. M. Savage, Born and Markov approximations for atom lasers, Phys. Rev. A \textbf{59}, 667 (1999).
\bibitem{ref-63} E. Knill, R. Laflamme, and G. J. Milburn, A scheme for efficient quantum computation with linear optics, Nature (London) \textbf{409}, 46 (2001).
\bibitem{ref-64} S. Rips, M. Kiffner, W.-Rae, and M. J. Hartmann, Steady-state negative Wigner functions of nonlinear nanomechanical oscillators, New J. Phys. \textbf{14}, 023042 (2012).
\bibitem{ref-65} Mohammad Hafezi and Peter Rabl, Optomechanically induced non-reciprocity in microring resonators, Opt. Express \textbf{20}, 7672-7684 (2012).
\bibitem{ref-Yang2505.10255} J. X. Yang, Cheng Shang, Yan-Hui Zhou, and H. Z. Shen, Simultaneous nonreciprocal unconventional photon blockade via two degenerate optical parametric amplifiers in spinning resonators, arXiv: 2505.10255 (2025).
\bibitem{ref-Yi2503.23169} H. Yi, T. Z. Luan, W. Y. Hu, Cheng Shang, Yan-Hui Zhou, Zhi-Cheng Shi, and H. Z. Shen, Nonreciprocity and unidirectional invisibility in three optical modes with non-Markovian effects, arXiv: 2503.23169 (2025).
\bibitem{ref-Luan2503.18647} T. Z. Luan, Cheng Shang, H. Yi, J. L. Li, Yan-Hui Zhou, Shuang Xu, and H. Z. Shen, Nonreciprocal quantum router with non-Markovian environments, arXiv: 2503.18647 (2025).
\bibitem{ref-66} H. J. Kimble, Strong interactions of single atoms and photons in cavity QED, Phys. Scr. \textbf{1998}, 127 (1998).
\bibitem{ref-67} J.-M. Pirkkalainen, S. U. Cho, Jian Li, G. S. Paraoanu, P. J. Hakonen, and M. A. Sillanp\"{a}\"{a}, Hybrid circuit cavity quantum electrodynamics with a mechanical resonator, Nature (London) \textbf{494}, 211 (2013).
\bibitem{ref-68} E. Verhagen, S. Del\'{e}glise, S. Weis, A. Schliesser, and T. J. Kippenberg, Quantum-coherent coupling of a mechanical oscillator to an optical cavity mode, Nature \textbf{482}, 63 (2012).
\bibitem{ref-69} J. D. Teufel, T. Donner, D. Li, J. W. Harlow, M. S. Allman, K. Cicak, A. J. Sirois, J. D. Whittaker, K. W. Lehnert, R. W. Simmonds, Sideband cooling of micromechanical motion to the quantum ground state, Nature (London) \textbf{475}, 359 (2011).
\bibitem{ref-70} J. C. Sankey, C. Yang, B. M. Zwickl, A. M. Jayich, and J. G. E. Harris, Strong and tunable nonlinear optomechanical coupling in a low-loss system, Nat. Phys. \textbf{6}, 707 (2010).
\bibitem{ref-71} J.-Q. Liao and F. Nori, Single-photon quadratic optomechanics, Scientific Reports \textbf{4}, 6302 (2014).
\bibitem{ref-exp-1} A. Nunnenkamp, V. Sudhir, A. K. Feofanov, A. Roulet, and T. J. Kippenberg, Quantum-Limited Amplification and Parametric Instability in the Reversed Dissipation Regime of Cavity Optomechanics, Phys. Rev. Lett. \textbf{113}, 023604 (2014).
\bibitem{ref-exp-2} D. Bothner, S. Yanai, A. Iniguez-Rabago, M. Yuan, Ya. M. Blanter, and G. A. Steele, Cavity electromechanics with parametric mechanical driving, Nat Commun \textbf{11}, 1589 (2020).
\bibitem{ref-72} E. M. Purcell, Spontaneous emission probabilities at radio frequencies, Phys. Rev. \textbf{69}, 681 (1946).
\bibitem{ref-73} J. M. Raimond, Brune, M. Haroche, S. Manipulating quantum entanglement with atoms and photons in a
cavity, Rev. Mod. Phys. \textbf{73}, 565 (2001).
\bibitem{ref-74} T. Niemczyk, F. Deppe, H. Huebl, E. P. Menzel, F. Hocke, M. J. Schwarz, J. J. Garcia-Ripoll, D. Zueco, T. H\"{u}mmer, E. Solano, A. Marx, R. Gross, Circuit quantum electrodynamics in the ultrastrong-coupling regime, Nature Physics \textbf{6}, 772 (2010).
\bibitem{ref-75} Andreas Bayer, Marcel Pozimski, Simon Schambeck, Dieter Schuh, Rupert Huber, Dominique Bougeard, and Christoph Lange, Terahertz light-matter interaction beyond unity coupling strength, Nano Lett. \textbf{17}, 6340 (2017).
\bibitem{ref-76} Anton Frisk Kockum, Adam Miranowicz, Simone De Liberato, Salvatore Savasta, and Franco Nori, Ultrastrong coupling between light and matter, Nature Reviews Physics \textbf{1}, 19-40 (2019).
\bibitem{ref-add-RH} Sh. Barzanjeh, M. H. Naderi, and M. Soltanolkotabi, Back-action ground-state cooling of a micromechanical membrane via intensity-dependent interaction, Phys. Rev. A \textbf{84}, 023803 (2011).
\bibitem{ref-77} E. X. DeJesus and C. Kaufman, Routh-Hurwitz criterion in the examination of eigenvalues of a system of nonlinear ordinary differential equations, Phys. Rev. A \textbf{35}, 5288 (1987).
\bibitem{ref-78} C. W. Gardiner and M. J. Collett, Input and output in damped quantum systems: Quantum stochastic differential equations and master equation, Phys. Rev. A \textbf{31}, 3761 (1985).
\bibitem{ref-81} X. W. Xu and Y. Li, Optical nonreciprocity and optomechanical circulator in three-mode optomechanical systems, Phys. Rev. A \textbf{91}, 053854 (2015).
\bibitem{ref-82} H. M\"{a}kel\"{a} and M. M\"{o}tt\"{o}nen, Effects of the rotating-wave and secular approximations on non-Markovianity, Phys. Rev. A \textbf{88}, 052111 (2013).
\bibitem{ref-83} Daniel Malz and Andreas Nunnenkamp, Optomechanical dual-beam backaction-evading measurement beyond the rotating-wave approximation, Phys. Rev. A \textbf{94}, 053820 (2016).
\bibitem{ref-84} L. Novotny and B. Hecht, \textit{Principles of Nano-Optics} (Cambridge University, UK, 2012).
\bibitem{ref-85} Cheng Shang, H. Z. Shen, and X. X. Yi, Nonreciprocity in a strong coupled three-mode optomechanical circulatory system, Optics Express \textbf{27}, 18 (2019).
\bibitem{ref-86} Mohammad-Ali Miri, Freek Ruesink, Ewold Verhagen, and Andrea Al\`{u}, Optical Nonreciprocity Based on Optomechanical Coupling, Phys. Rev. Applied \textbf{7}, 064014 (2017).
\bibitem{ref-87} Yi-Bing Qian, Deng-Gao Lai, Mei-Ran Chen, and Bang-Pin Hou, Nonreciprocal photon transmission with quantum noise reduction via cross-Kerr nonlinearity, Phys. Rev. A \textbf{104}, 033705 (2021).
\bibitem{ref-92} G. S. Agrwal and S. Huang, Electromagnetically induced transparency in mechanical effects of light, Phys. Rev. A \textbf{81}, 041803(R) (2010).
\bibitem{ref-93} S. Weis, R. Rivi\`{e}re, S. Del\'{e}glise, E. Gavartin, O. Arcizet, A. Schliesser, and T. J. Kippenberg, Optomechanically induced transparency, Science \textbf{330}, 1520 (2010).
\bibitem{ref-94} A. H. Safavinaeini, T. P. M. Alegre, J. Chan, M. Eichenfield, W. Winger, Q. Lin, J. T. Hill, D. E. Chang, and O. Painter, Electromagnetically induced transparency and slow light with optomechanics, Nature(London) \textbf{472}, 69 (2011).
\bibitem{ref-95} D. Vitali, S. Gigan, A. Ferreira, H. R. B\"{o}hm, P. Tombesi, A. Guerreiro, V. Vedral, A. Zeilinger, and M. Aspelmeyer, Optomechanical Entanglement between a Movable Mirror and a Cavity Field, Phys. Rev. Lett. \textbf{98}, 030405 (2007)
\bibitem{ref-96} Y.-F. Jiao, S.-D. Zhang, Y.-L. Zhang, A. Miranowicz, L.-M. Kuang, and H. Jing, Nonreciprocal Optomechanical Entanglement against Backscattering Losses, Phys. Rev. Lett. \textbf{125}, 143605 (2020).
\bibitem{ref-97} S. Barzanjeh, A. Xuereb, S. Gr\"{o}blacher,  M. Paternostro, C. A. Regalet, and Eva M. Weig, Optomechanics for quantum technologies, Nature Physics, \textbf{18}, 15-24 (2022).
\bibitem{ref-Shang044048} Cheng Shang and Hongchao Li, Resonance-dominant optomechanical entanglement in open quantum systems, Phys. Rev. Applied \textbf{21}, 044048 (2024).
\bibitem{ref-Sun2504.09695} J. Y. Sun, C. Cui, Y. F. Li, Shuang Xu, Cheng Shang, Yan-Hui Zhou, and H. Z. Shen, Dressed bound states and non-Markovian dynamics with a whispering-gallery-mode microcavity coupled to a two-level atom and a semi-infinite photonic waveguide, arXiv: 2504.09695 (2025).
\bibitem{ref-Shen2503.21739} H. Z. Shen, Cheng Shang, Yan-Hui Zhou, and X. X. Yi, Emergent Non-Markovian Gain in Open Quantum Systems, arXiv: 2503.21739 (2025).
\bibitem{ref-Zhang063702} Bo-Wang Zhang, Cheng Shang, J. Y. Sun, Zhuo-Cheng Gu, and X. X. Yi, Manipulating spectral transitions and photonic transmission in a non-Hermitian optical system through nanoparticle perturbations, Phys. Rev. A \textbf{111}, 063702 (2025).
\bibitem{ref-Ning250400617} L. Y. Ning, Zhi-Guang Lu, Cheng Shang, and H. Z. Shen, Higher-order Exceptional Points Induced by Non-Markovian Environments, arXiv: 2504.00617 (2025).
\bibitem{ref-98} M.-A. Miri, Freek Ruesink, Ewold Verhagen, and Andrea Al\`{u}, Optical nonreciprocity based on optomechanical coupling, Phys. Rev. Applied \textbf{7}, 064014 (2017).
\bibitem{ref-99} F. Ruesink, M.-A. Miri, Andrea Al\`{u}, Ewold Verhagen, Nonreciprocity and magnetic-free isolation based on optomechanical interactions, Nature communications \textbf{7}, 1-8 (2016).
\end{references}
\end{document}